\documentclass[journal]{ieeetran}

\usepackage{cite}
\usepackage{amsmath,amssymb,amsfonts}
\usepackage{algorithmic}
\usepackage{graphicx}
\usepackage{float}
\usepackage{textcomp}
\def\BibTeX{{\rm B\kern-.05em{\sc i\kern-.025em b}\kern-.08emT\kern-.1667em\lower.7ex\hbox{E}\kern-.125emX}}
\usepackage{url}
\newtheorem{thm}{Theorem}
\newtheorem{defn}{Definition}

\newtheorem{lem}{Lemma}
\newtheorem{prob}{Problem}

\newtheorem{assm}{Assumption}
\newtheorem{reform}{Reformulation}

\newcommand{\R}{\mathbb{R}}
\newcommand{\N}{\mathbb{N}}
\renewcommand{\P}{\mathbb{P}}
\newcommand{\E}{\mathbb{E}}
\newcommand{\U}{\mathcal{U}}
\newcommand{\X}{\mathcal{X}}
\newcommand{\T}{\mathcal{T}}

\newcommand{\pr}[1]{\mathbb{P}\!\left(#1\right)}
\newcommand{\ex}[1]{\mathbb{E}\!\left[#1\right]}
\newcommand{\hex}[1]{\hat{\mathbb{E}}\!\left[#1\right]}

\newcommand{\var}[1]{\mathrm{Var}\!\left(#1\right)}
\newcommand{\hvar}[1]{\hat{\mathrm{Var}}\!\left(#1\right)}
\newcommand{\std}[1]{\mathrm{Std}\!\left(#1\right)}
\newcommand{\hstd}[1]{\hat{\mathrm{Std}}\!\left(#1\right)}
\newcommand{\cov}[2]{\mathrm{Cov}\!\left(#1,#2\right)}
\newcommand{\hcov}[2]{\hat{\mathrm{Cov}}\!\left(#1,#2\right)}
\newcommand{\bvec}[1]{\vec{\boldsymbol{#1}}}
\newcommand{\Nt}[2]{\mathbb{N}_{[#1,#2]}}

\makeatletter
\let\NAT@parse\undefined
\makeatother
\usepackage{hyperref}
\hypersetup{
    colorlinks=true,      
    linkcolor=black,
    citecolor=black,
    filecolor=black,
    urlcolor=black     
    }

\author{Shawn Priore, \IEEEmembership{Student Member, IEEE}, and Meeko Oishi, \IEEEmembership{Senior Member, IEEE}
\thanks{This material is based upon work supported by the National Science Foundation under NSF Grant Number CMMI-2105631. (Corresponding author: Shawn Priore)} 
\thanks{Shawn Priore, and Meeko Oishi are with the Department of Electrical and Computer Engineering, University of New Mexico, Albuquerque, NM 87131 (e-mail: \texttt{shawn.a.priore@gmail.com, oishi@unm.edu}).
    }
}

\title{Chance Constrained Stochastic Optimal Control Based on Sample Statistics With Almost Surely Probabilistic Guarantees}

\begin{document}

\maketitle

\begin{abstract}
While techniques have been developed for chance constrained stochastic optimal control using sample disturbance data that provide a probabilistic confidence bound for chance constraint satisfaction, far less is known about how to use sample data in a manner that can provide almost surely guarantees of chance constraint satisfaction. In this paper, we develop a method for solving chance constrained stochastic optimal control problems for linear time-invariant systems with unknown but sampled additive disturbances. We propose an open-loop control scheme for multi-vehicle planning, with both target sets and collision avoidance constraints. To this end, we derive a concentration inequality to bound the tail probability of a random variable based on sample statistics. Using the derived concentration inequality, we reformulate each chance constraint in terms of the sample mean and sample standard deviation. While the reformulated bounds are conservative with respect to the original bounds, they have a simple and closed form, guarantee chance constraint satisfaction almost surely, and are amenable to difference of convex optimization techniques. We demonstrate our approach on two multi-satellite rendezvous problems. 
\end{abstract}


\section{Introduction}

Safety critical systems, such as satellites, aircraft, and autonomous vehicles, often require assurances that synthesized controllers will not lead to unsafe or potentially detrimental trajectories. However, when dynamics are corrupted by stochasticity from modeling inaccuracies or external forces, providing such assurances can be challenging as low probability events can result in dangerous trajectories. In scenarios such as these, assurances typically can only be made probabilistically with a desired level of risk. For stochastic processes with a known form, such as additive Gaussian noise, stochastic optimal control techniques can be used to provide such assurances. However, when the underlying stochastic process is unknown, many of these techniques cannot be used as they require analytic forms of the constraint. In the place of an analytic form, collected disturbance samples can be used to provide an approximate or asymptotic probabilistic assurance provided a large enough sample size has been collected. However, while several techniques exist to synthesize controllers with sample data, few methods exist that are both computationally efficient and can provide almost surely assurances of probabilistic constraint satisfaction.

% One of the primary challenges preventing sample based methods from providing assurances of probabilistic constraint satisfaction, is lack of knowledge of the underlying process. Without knowledge of the underlying distribution, analytic techniques cannot be evoked to evaluate the chance constraint probability. 

Particle control approaches utilize sample data to synthesize a controller that satisfies the constraint for a percentage of samples corresponding to the probabilistic safety threshold to avoid the need for analytic evaluations \cite{blackmore2010_particle}. However, the particle control approach does not have a finite sample confidence bound and can only provide satisfaction assurances asymptotically \cite{Blackmore2006}. In contrast, the scenario approach relies on synthesizing a control that satisfies the constraint for each disturbance sample in the sample set \cite{calafiore2006scenario, Campi2008}. For finite sample sizes, the scenario approach guarantees the synthesized controller satisfies the chance constraint and is optimal up to a probabilistic confidence bound \cite{calafiore2006scenario}. While the scenario approach, and similarly the particle control approach, suffer from computational burden of large sample sizes, methods have been proposed to decrease the computational burden by discarding samples \cite{campi2011sampling}, as well as optimizing over a subset of the samples \cite{care2014fast}. 

In contrast to approaches which employ the entire sample set, data reduction methods have been posed to simplify sample-based approaches based on parameter estimation \cite{Saha2010, Madankan15, RAJAMANI09}, a common tool from statistical literature. By computing moments or extremum of the sample set, probabilistic evaluations can be computed via robust control principals \cite{Ben-Tal2009, Lam15} or taken as ground truth and incorporated into moment based approaches \cite{Verma10}. While robust and ground truth assumptions can be computationally efficient and sufficient asymptotically via the central limit theorem, computation with finite sample sizes does not provide any probabilistic assurances of chance constraint satisfaction.

We propose an approach that relies on data reduction via moment estimation, yet allows for {\em almost surely} probabilistic guarantees of chance constraint satisfaction, within a difference-of-convex programming framework. Our approach handles completely arbitrary disturbances, even ones for which moments do not exist analytically, because it relies upon sample moment evaluation, which exists for any random variable. Further, data reduction via moment estimation allows for more efficient computation than particle and scenario based approaches. We apply our approach to chance constraint evaluation that arises in multi-vehicle planning problems: that is, in a) reaching a terminal target set and b) avoiding collision with obstacles in the environment as well as with other vehicles. 

The main contribution of this paper is a {\em closed-form} reformulation of chance constraints based on sample statistics, under arbitrary distributions, for polytopic target sets and collision avoidance constraints, that is amenable to difference of convex programming solutions. Because this approach relies upon sample estimations of the first and second moments only, it inherently is limited in its ability to capture modality, skewness or other high order moment phenomena.  Although it is conservative with respect to other sample based methods, we have found the conservatism to be reasonable, in practice.  

The paper is organized as follows. Section \ref{sec:prelim} provides mathematical preliminaries and formulates the optimization problem. Section \ref{sec:methods} derives a concentration inequality based on sample statistics and derives the difference of convex functions optimization problem reformulation of the chance constraints. Section \ref{sec:results} demonstrates our approach on a multi-satellite rendezvous problem, and Section \ref{sec:conclusion} provides concluding remarks.

\section{Preliminary and Problem Setup} \label{sec:prelim}

\subsection{Mathematical Preliminaries}

We denote the interval that enumerates all natural numbers from $a$ to $b$, inclusively, as $\Nt{a}{b}$. We denote vectors with an arrow accent, as $\vec{x} \in \R^n$. Random variables are indicated with a bold case $\boldsymbol{x}$. For a random variable $\boldsymbol{x}$, we denote the expectation as $\ex{\boldsymbol{x}}$, variance as $\var{\boldsymbol{x}}$, and standard deviation as $\std{\boldsymbol{x}}$. For a vector input, $\var{\cdot}$ will reference the variance-covariance matrix of the random vector. For two random variables, $\boldsymbol{x}$ and $\boldsymbol{y}$, $\cov{\boldsymbol{x}}{\boldsymbol{y}}$ denotes the covariance between the two variables. Samples will be denoted with bracketed superscript, as $\cdot^{[j]}$, and sample estimates will have a hat accent $\hat{\cdot}$. We denote the 2-norm of a matrix or vector by $\|\cdot\|$. The $n \times n$ identity matrix is $I_n$, a $n \times m$ matrix of zeros is $0_{n \times m}$ or $0_{n}$ if $n=m$, and a $n\times 1$ vector of ones is $\vec{1}_n$.

\subsection{Problem Formulation}

\begin{figure}
    \centering
    \includegraphics[width=0.7\columnwidth]{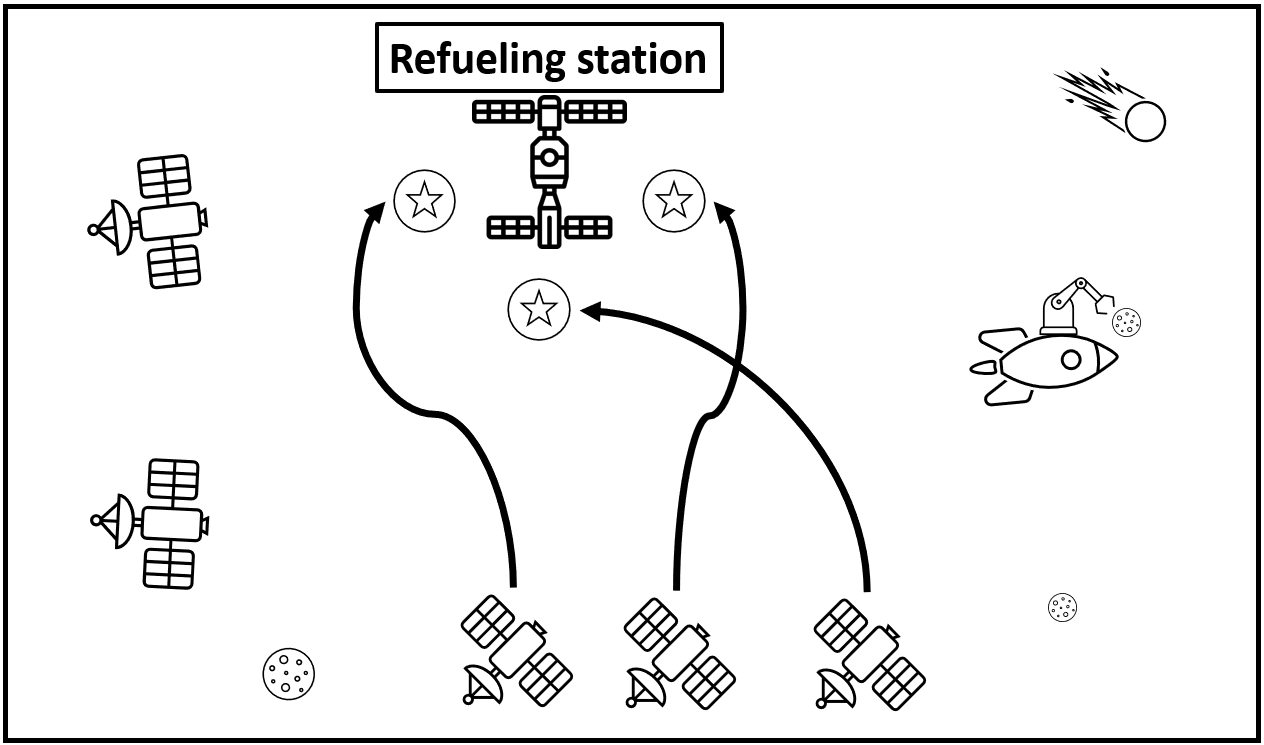}
    \caption{A scenario in which three satellites need to rendezvous with a refueling station while avoiding nearby spacecraft and space debris.}
    \label{fig:demo}
\end{figure}

Consider a scenario, such as the one shown in Figure \ref{fig:demo}, in which three satellites rendezvous with a refueling station while avoiding each other, other spacecraft, and debris. With arbitrary and unknown but sampled disturbances corrupting the satellite dynamics, we seek to synthesize a controller to construct an optimal rendezvous maneuver that satisfies chance constraints target set and collision avoidance.

We presume the evolution of $N_{v}$ vehicles are governed by the discrete-time LTI system,
\begin{equation}
    \bvec{x}_i(k+1) = A \bvec{x}_i(k) + B \vec{u}_i(k) + \bvec{w}_i(k) \label{eq:system}
\end{equation}
with state $\bvec{x}_i(k) \in \X \subseteq \R^n$, input $\vec{u}_i(k) \in \U \subseteq \R^m$,  disturbance $\bvec{w}_i(k) \in \R^n$, and known initial condition $\vec{x}_i(0)$. We presume the admissible control set $\U$ is a closed, compact, and convex polytope. The system evolves over a finite time horizon of $N \in \N$ steps. We presume each disturbance, $\bvec{w}_i(k)$, is continuous and has probability space $(\Omega, \mathcal{B}(\Omega), \P_{\bvec{w}_i (k)})$ with outcomes $\Omega$, Borel $\sigma$-algebra $\mathcal{B}(\Omega)$, and probability measure $\P_{\bvec{w}_i (k)}$ \cite{casella2002}. Further, the disturbance is considered arbitrary and unknown.

We write the dynamics at time $k$ as an affine combination of the initial condition and the concatenated control sequence and concatenated disturbance,
\begin{equation} \label{eq:lin_dynamics}
    \bvec{x}_i(k) = A^k \vec{x}_i(0) + \mathcal{C}(k) \vec{U}_i + \mathcal{D}(k) \bvec{W}_i
\end{equation}
with 
\begin{subequations}
\begin{align}
    \vec{U}_i =& \left[ \vec{u}_i(0)^\top \ \ldots \ \vec{u}_i(N-1)^\top \right]^\top &&\in \mathcal{U}^{N} \\
    \bvec{W}_i =& \left[ \bvec{w}_i(0)^\top \ \ldots \ \bvec{w}_i(N-1)^\top \right]^\top &&\in \mathbb{R}^{Nn} \\
    \mathcal{C}(k) = & \left[ A^{k-1}B \ \ldots \ AB \ B \ 0_{n \times (N\!-\!k)m} \right] && \in \R^{n \times Nm} \\
    \mathcal{D}(k) = & \left[ A^{k-1} \ \ldots \ A \ I_n \ 0_{n \times (N\!-\!k)n} \right] && \in \R^{n \times Nn}
\end{align}
\end{subequations}    

We presume a convex performance objective $J: \X^{N \times N_{v}} \times \U^{N \times N_{v}} \rightarrow \R$, such as fuel cost. We presume desired time-varying polytopic target sets that each satellite must reach, obstacles with known positions that it must avoid, as well as the need for collision avoidance between vehicles, all with desired likelihoods,
\begin{subequations}\label{eq:constraints}
    \begin{align}
    \pr{\bigcap_{i=1}^{N_{v}}  \bigcap_{k=1}^{N} \bvec{x}_i(k)  \in  \T_{i}(k)} & \geq  1\!-\!\alpha \label{eq:constraint_t}\\
    \pr{\bigcap_{i=1}^{N_{v}}  \bigcap_{k=1}^{N} \| S (\bvec{x}_i(k) \!-\! \vec{o}(k)) \| \geq r} &\geq 1\!-\!\beta \label{eq:constraint_o}\\
    \pr{ \bigcap_{i=1}^{N_v-1} \bigcap_{j=i+1}^{N_v}  \bigcap_{k=1}^{N} \| S (\bvec{x}_i(k) \!-\!  \bvec{x}_j (k)) \| \geq r } &\geq 1\!-\!\gamma \label{eq:constraint_r}
    \end{align}
\end{subequations}
We presume convex, compact, and polytopic sets $ \T_{i}(k) \subseteq \R^n$, matrix $S \in \mathbb R^{q \times n}$, positive scalar $r \in \mathbb R_+$, non-random object locations $\vec{o}(k) \in \R^n$, and probabilistic violation thresholds $\alpha, \beta, \gamma$. Here, $S$ is designed to extract the position of the vehicle from the state vector.

\begin{defn}[Reverse convex constraint] \label{def:reverse-convex}
The complement of a convex constraint is a reverse convex constraint. That is, for a convex function $f: \R \rightarrow \R$ and a scalar $c \in \R$, a reverse convex constraint has the form $f(x) \geq c$.
\end{defn}

Note that the collision avoidance constraints inside the probability functions are reverse-convex as per Definition \ref{def:reverse-convex}. 

We seek to solve the following optimization problem.
\begin{subequations}\label{eq:prob_1_opt}
    \begin{align}
        \underset{\vec{U}_1, \dots, \vec{U}_{N_{v}}}{\mathrm{minimize}} \quad & J\left(
        \bvec{X}_1, \ldots, \bvec{X}_{N_{v}},  \vec{U}_1, \dots, \vec{U}_{N_{v}}\right)  \\
        \mathrm{subject\ to} \quad  & \vec{U}_1, \dots, \vec{U}_{N_{v}} \in  \mathcal U^N,  \\
        & \text{Dynamics } \eqref{eq:lin_dynamics} \text{ with }
        \vec{x}_1(0), \dots, \vec{x}_{N_{v}}(0)\\
        & \text{Probabilistic constraints  \eqref{eq:constraints}} \label{eq:prob_1_opt_constraints} 
    \end{align}
\end{subequations}
where $\bvec{X}_i = \begin{bmatrix} \bvec{x}_i^{\top}(1) & \ldots & \bvec{x}_i^{\top}(N) \end{bmatrix}^{\top}$ is the concatenated state vector for vehicle $i$. 

\begin{defn}[Almost Surely \cite{casella2002}]
    Let $(\Psi, \mathcal{B}(\Psi), \P)$ be a probability space with outcomes $\Psi$, Borel $\sigma$-algebra $\mathcal{B}(\Psi)$, and probability measure $\P$. An event $\mathcal{A} \in \mathcal{B}(\Psi)$ happens almost surely if $\pr{\mathcal{A}} = 1$ or $\pr{\mathcal{A}^{c}}=0$ where $\cdot^{c}$ denotes the complement of the event.
\end{defn}

As the disturbance is unknown and arbitrary, we make several key assumptions about the quantity and quality of the sampled disturbance data that allow us to make \eqref{eq:prob_1_opt} a tractable problem.

\begin{assm} \label{assm:samples}
For each vehicle, the concatenated disturbance vector $\bvec{W}_i$ has been independently sampled $N_s$ times. We denote the sampled values as $\bvec{W}_i^{[j]}$ for $j \in \Nt{1}{N_s}$. 
\end{assm}
\begin{assm} \label{assm:n_samples}
The sample size $N_s$ must be sufficiently large such that the reformulations presented in this work are tractable.
\end{assm}
\begin{assm} \label{assm:samples_not_equal}
For each vehicle, the concatenated disturbance vector samples $\bvec{W}_i^{[j]}$ are almost surely not all equal.
\end{assm}

Assumptions \ref{assm:samples}-\ref{assm:n_samples} are required to compute sample mean and standard deviation. Assumption \ref{assm:n_samples} guarantees that the sample based concentration inequality developed here can be applied for our reformulations. Assumption \ref{assm:samples_not_equal} guarantees the distribution is not degenerate or deterministic. 

\begin{prob} \label{prob:1}
    Under Assumptions \ref{assm:samples}-\ref{assm:samples_not_equal}, solve the stochastic optimization problem \eqref{eq:prob_1_opt} with probabilistic violation thresholds $\alpha$, $\beta$, and $\gamma$ for open loop controllers $\vec{U}_1, \dots, \vec{U}_{N_v} \in  \mathcal U^N$.
\end{prob} 

The main challenge in solving Problem \ref{prob:1} is assuring \eqref{eq:prob_1_opt_constraints}. Without knowledge of the underlying distribution, analytic techniques cannot be used to derive reformulations that allow for guarantees. Further, current sample based methods can only guarantee chance constraint satisfaction approximately or asymptotically.

\section{Methods} \label{sec:methods}

Our approach to solve Problem \ref{prob:1} involves  reformulating each chance constraint as an affine summation of the random variable's sample mean and sample standard deviation, i.e., $\hex{\| S (\bvec{x}_i(k) \!-\!  \bvec{x}_j (k)) \|}$ and $\hstd{\| S (\bvec{x}_i(k) \!-\!  \bvec{x}_j (k)) \|}$, respectively for the collision avoidance constraint. The reformulation is a result of a concentration inequality based on sample statistics that is developed in this work. Said concentration inequality guarantees satisfaction of \eqref{eq:constraints} almost surely. 

\subsection{Establishing Sample Bounds}

Here, we state the pivotal theorem that allow us to solve Problem \ref{prob:1}. For brevity, the proof is in Appendix \ref{appx:out-sample}. 

\begin{thm} \label{thm:out_of_sample}
Let $\boldsymbol{x}$ follow some distribution $f$. Let $\boldsymbol{x}^{[1]},\dots, \boldsymbol{x}^{[N_s]}$ be samples drawn independently from the distribution $f$, for some $N_s\geq2$. Let
\begin{subequations} \label{eq:stats}
\begin{align}
    \hex{\boldsymbol{x}} = & \; \frac{1}{N_s} \sum_{i=1}^{N_s} \boldsymbol{x}^{[i]} \\
    \hstd{\boldsymbol{x}} =  & \; \sqrt{\frac{1}{N_s} \sum_{i=1}^{N_s} (\boldsymbol{x}^{[i]} - \hex{\boldsymbol{x}})^2}
\end{align}
\end{subequations}
be the sample mean and sample standard deviation, respectively, with $\hstd{\boldsymbol{x}}>0$ almost surely. Then for any $\lambda>0$
\begin{equation} \label{eq:out_sample_cantelli}
    \pr{\boldsymbol{x}-\hex{\boldsymbol{x}} \geq \lambda \hstd{\boldsymbol{x}}} \leq \frac{(\sqrt{N_s + 1} + \lambda)^2}{\lambda^2 N_s + (\sqrt{N_s + 1}+\lambda)^2 }
\end{equation}
\end{thm}

Theorem \ref{thm:out_of_sample} provides a bound for deviations of a random variable $\boldsymbol{x}$ from the sample mean. For the purpose of generating open loop controllers in an optimization framework, this will allow us to bound chance constraints based on sample statistics of sample disturbance data.

For brevity, we define $N_s^{\ast} = N_S + 1$ and 
\begin{equation} \label{eq:lambda_func}
    f(\lambda) = \frac{(\sqrt{N_s^{\ast}} + \lambda)^2}{\lambda^2 N_s + (\sqrt{N_s^{\ast}} + \lambda)^2} 
\end{equation}

\begin{figure}
    \centering
    \includegraphics[width=0.9\columnwidth]{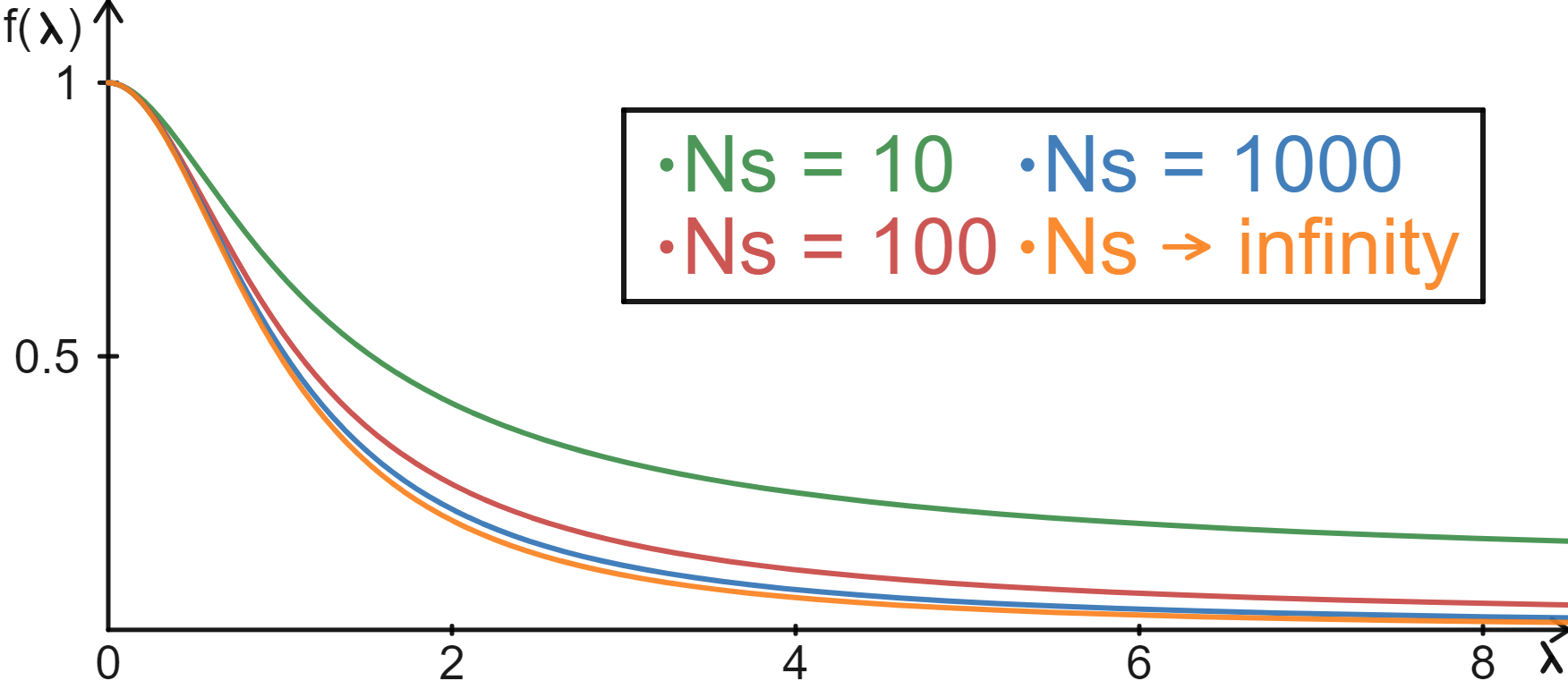}
    \caption{Graph of \eqref{eq:f_lim} for values of $N_s \in \{10,100,1000\}$ and as $N_s \rightarrow \infty$. }
    \label{fig:ineq}
\end{figure}

To address the need for Assumption \ref{assm:n_samples}, we observe that
\begin{equation} \label{eq:f_lim}
    \lim_{\lambda \rightarrow \infty} f(\lambda) = \frac{1}{N_s^{\ast}}
\end{equation}
For any probabilistic violation threshold smaller than this value, Theorem \ref{thm:out_of_sample} will not be sufficiently tight to bound the constraint. Figure \ref{fig:ineq} graphs \eqref{eq:lambda_func} for $N_s$ taking the values 10, 100, and 1000, and as $N_s \rightarrow \infty$.

Note that we did not use Bessel's correction \cite{casella2002} in the sample variance formula to simplify the probabilistic bound. Accordingly, the sample variance statistic we used is biased in relation to the variance of the distribution. An analogous result can be derived with Bessel's correction, however, the bound becomes more complex.

We can easily find the lower tail bound with Theorem \ref{thm:out_of_sample} by substituting $\boldsymbol{x}$ with $-\boldsymbol{x}$ as
\begin{equation} 
    \pr{\boldsymbol{x} - \hex{\boldsymbol{x}} \leq -\lambda \hstd{\boldsymbol{x}}} \leq f(\lambda)
\end{equation}
This observation will be useful in the reformulation of the collision avoidance constraints.

\subsection{Polytopic Target Set Constraint} \label{ssec:target_reform}

First, consider the reformulation of \eqref{eq:constraint_t}. Without loss of generality, we presume $N_v=1$ and $N=1$ for brevity. The polytope $\T_{i}(k)$ can be written as the intersection of $N_{Tik}$ half-space inequalities,
\begin{equation} \label{eq:polytope}
    \pr{\bvec{x}_i(k) \in \T_{i}(k)} =   \pr{\bigcap_{j=1}^{N_{Tik}} \vec{G}_{ijk} \bvec{x}_i(k) \leq h_{ijk}}
\end{equation}
where $\vec{G}_{ijk} \in \R^n$ and $h_{ijk} \in \R$. We take the complement and employ Boole's inequality to separate the combined chance constraints into a series of individual chance constraints,
\begin{subequations}
\begin{align}
    \pr{\boldsymbol{x}_i(k) \not\in \T_i (k)} = & \; \pr{ \bigcup_{j=1}^{N_{Tik}} \vec{G}_{ijk}\bvec{x}_i(k) \geq h_{ijk}} \\
    \leq & \; \sum_{j=1}^{N_{Tik}} \pr{ \vec{G}_{ijk}\bvec{x}_i(k) \geq h_{ijk}}
\end{align}
\end{subequations}
Using the approach in \cite{ono2008iterative}, we introduce $\omega_{ijk}$ to allocate risk for each individual chance constraint,
\begin{subequations}\label{eq:quantile_reform_new_var}
\begin{align}
     \pr{ \vec{G}_{ijk}\bvec{x}_i(k) \geq h_{ijk}} &\leq \omega_{ijk} \label{eq:quantile_orig} \\
     \sum_{j=1}^{N_{Tik}} \omega_{ijk} &\leq \alpha \label{eq:quantile_reform_new_var_2}\\
     \omega_{ijk} & \geq 0 \label{eq:quantile_reform_new_var_3}
\end{align}
\end{subequations}
To find a solution to \eqref{eq:quantile_reform_new_var}, we need to find an appropriate value for $\omega_{ijk}$. To that end, we add the additional constraints 
\begin{subequations} \label{eq:add_target}
\begin{align}
     \hex{\vec{G}_{ijk}\bvec{x}_i(k)} + \lambda_{ijk} \hstd{\vec{G}_{ijk}\bvec{x}_i(k)} & \; \leq h_{ijk} \\
     \lambda_{ijk} & \; > 0
\end{align}
\end{subequations}
with $\lambda_{ijk}>0$ where $\hat{\E}[\vec{G}_{ijk}\bvec{x}_i(k)]$ and $\hat{\mathrm{Std}}(\vec{G}_{ijk}\bvec{x}_i(k))$ are the sample mean and sample standard deviation, respectively. For brevity, we denote
\begin{equation}
    \hat{\mathbb{F}}(\bvec{x}_i(k), \lambda_{ijk}) =\hex{\vec{G}_{ijk}\bvec{x}_i(k)} + \lambda_{ijk} \hstd{\vec{G}_{ijk}\bvec{x}_i(k)}
\end{equation}
Enforcement of \eqref{eq:add_target} guarantees that  
\begin{equation} \label{eq:enforce_target}
    \pr{ \vec{G}_{ijk}\bvec{x}_i(k) \geq h_{ijk}} \! \leq \pr{ \vec{G}_{ijk}\bvec{x}_i(k) \geq \hat{\mathbb{F}}(\bvec{x}_i(k), \lambda_{ijk}) }
\end{equation} 
allowing us to write \eqref{eq:quantile_reform_new_var} as
\begin{subequations}\label{eq:quantile_reform_new_var_1.5}
\begin{align}
     \pr{ \vec{G}_{ijk}\bvec{x}_i(k) \geq \hat{\mathbb{F}}(\bvec{x}_i(k), \lambda_{ijk}) } &\leq \omega_{ijk}  \\
     \hat{\mathbb{F}}(\bvec{x}_i(k), \lambda_{ijk}) & \geq h_{ijk} \\
     \sum_{j=1}^{N_{Tik}} \omega_{ijk} &\leq \alpha \\
     \omega_{ijk} & \geq 0 \\
     \lambda_{ijk} & > 0
\end{align}
\end{subequations}
Then, by Assumptions \ref{assm:samples}-\ref{assm:samples_not_equal} and Theorem \ref{thm:out_of_sample}, we can substitute
\begin{equation} \label{eq:target_omega}
    \omega_{ijk} = f(\lambda_{ijk})
\end{equation}
via \eqref{eq:out_sample_cantelli}. Then \eqref{eq:quantile_reform_new_var_1.5} becomes
\begin{subequations}\label{eq:reform_2}
\begin{align}
     \pr{ \vec{G}_{ijk}\bvec{x}_i(k) \geq \hat{\mathbb{F}}(\bvec{x}_i(k), \lambda_{ijk}) } & \leq f(\lambda_{ijk})  \label{eq:reform_2_1}\\
     \hat{\mathbb{F}}(\bvec{x}_i(k), \lambda_{ijk}) & \geq h_{ijk} \label{eq:reform_2_2}\\
     \sum_{j=1}^{N_{Tik}} f(\lambda_{ijk}) &\leq \alpha \label{eq:reform_2_3}\\
     f(\lambda_{ijk}) & \geq 0 \label{eq:reform_2_4} \\
     \lambda_{ijk} & > 0 \label{eq:reform_2_5}
\end{align}
\end{subequations}

To use Theorem \ref{thm:out_of_sample}, we must impose the restriction $N_s \geq 2$. This restriction will be in addition to Assumption \ref{assm:n_samples}.

We can simplify \eqref{eq:quantile_reform_new_var_2} as \eqref{eq:reform_2_2}-\eqref{eq:reform_2_3} enforce \eqref{eq:reform_2_1} and \eqref{eq:reform_2_5} enforces \eqref{eq:reform_2_4}. Hence, \eqref{eq:quantile_reform_new_var_2} becomes
\begin{subequations}\label{eq:target_constraint}
\begin{align}
     \hat{\mathbb{F}}(\bvec{x}_i(k), \lambda_{ijk}) & \geq h_{ijk} \label{eq:target_reform}\\
     \sum_{j=1}^{N_{Tik}} f(\lambda_{ijk}) &\leq \alpha \label{eq:target_lambda}\\
     \lambda_{ijk} & > 0 \label{eq:lambda_restrict}
\end{align}
\end{subequations}
with \eqref{eq:target_reform} and \eqref{eq:lambda_restrict} iterated over the index $j \in \Nt{1}{N_{Tik}}$.

Next, we show that $\hat{\mathbb{F}}(\bvec{x}_i(k), \lambda_{ijk})$ has a closed and convex form. Observe that the sample mean vector and sample variance-covariance matrix of $\bvec{W}$ are
\begin{subequations}
\begin{align}
    \hex{\bvec{W}_i} & = \frac{1}{N_s} \sum_{[i]=1}^{N_s} \bvec{W}_i^{[i]} \\
    \hvar{\bvec{W}_i} & =  \frac{1}{N_s}  \sum_{[i]=1}^{N_s} \left(\bvec{W}_i^{[i]} \!-\!\hex{\bvec{W}_i}\right)\left(\bvec{W}_i^{[i]} \!-\!\hex{\bvec{W}_i}\right)^{\top}
\end{align}
\end{subequations}
Then, the sample mean and standard deviation for the each half-space constraint is 
\begin{subequations} \label{eq:sample_poly_mean_var}
\begin{align}
    \hex{\vec{G}_{ijk}\bvec{x}_i(k)} & = \vec{G}_{ijk} \left(A^k \vec{x}_i(0) \!+\! \mathcal{C}(k) \vec{U}_i \!+\! \mathcal{D}(k) \hex{\bvec{W}_i} \right) \label{eq:sample_poly_mean} \\
    \hstd{\vec{G}_{ijk}\bvec{x}_i(k)} & = \sqrt{\vec{G}_{ijk}\mathcal{D}(k) \hvar{\bvec{W}_i} \mathcal{D}^{\top}(k)\vec{G}^{\top}_{ijk}} \label{eq:sample_poly_var}
\end{align}
\end{subequations}
So,
\begin{equation}
    \begin{split}
        \hat{\mathbb{F}}&(\bvec{x}_i(k), \lambda_{ijk}) \\
        & = \vec{G}_{ijk} \left(A^k \vec{x}_i(0) \!+\! \mathcal{C}(k) \vec{U}_i \!+\! \mathcal{D}(k) \hex{\bvec{W}_i} \right) \\
        & \qquad + \lambda_{ijk} \sqrt{\vec{G}_{ijk}\mathcal{D}(k) \hvar{\bvec{W}_i} \mathcal{D}^{\top}(k)\vec{G}^{\top}} 
    \end{split}
\end{equation}
which is affine, and hence convex, in the control input.

\begin{figure*}
\setcounter{equation}{27}
\begin{subequations}\label{eq:target_constraint_3}
\begin{align}
    \vec{G}_{ijk}\left( A^k \vec{x}_i(0) + \mathcal{C}(k) \vec{U}_i + \mathcal{D}(k) \hex{\bvec{W}_i} \right) + \lambda_{ijk} \sqrt{\vec{G}_{ijk}\mathcal{D}(k) \hvar{\bvec{W}_i} \mathcal{D}^{\top}(k)\vec{G}^{\top}_{ijk}} \leq & \; h_{ijk} & \forall j \in \Nt{1}{N_{Tik}} \label{eq:target_reform_3}\\ 
    \sum_{j=1}^{N_{Tik}} \frac{(\sqrt{N_s^{\ast}}+\lambda_{ijk})^2}{\lambda_{ijk}^2 N_s+(\sqrt{N_s^{\ast}}+\lambda_{ijk})^2 } \leq & \; \alpha & \label{eq:target_lambda_3}\\
    \sqrt{N_s^{\ast}}\left[\cos\left(\frac{1}{3}\arccos\left(-\frac{N_s-1}{N_s^{\ast}}\right)\right)-\frac{1}{2}\right] \leq & \;  \lambda_{ijk} & \forall j \in \Nt{1}{N_{Tik}} \label{eq:lambda_restrict_3}
\end{align}
\end{subequations}
\hrulefill
\setcounter{equation}{23}
\end{figure*}

Finally, we note that \eqref{eq:target_lambda} is not convex over the set $\lambda_{ijk} > 0$. Here, we must find the values of $\lambda_{ijk}$ that make the constraint \eqref{eq:target_lambda} convex. Observe the second partial derivative of $f(\lambda)$ with respect to $\lambda$ is
\begin{equation}
    \frac{\partial^2}{\partial \lambda^2}f(\lambda) = \frac{2N_s\left(\lambda^3(N_s^{\ast})^{3/2}+3\lambda^2(N_s^{\ast})^2-(N_s^{\ast})^2\right)}{\left(N_s\lambda^2+\left(\lambda+\sqrt{N_s^{\ast}}\right)^2\right)^3}    
\end{equation}
Then $f(\lambda)$ has inflection points where 
\begin{equation} \label{eq:cubic_eq}
    \frac{2}{\sqrt{N_s^{\ast}}}\lambda^3+3\lambda^2 -1 = 0
\end{equation}
The function \eqref{eq:cubic_eq} has three real roots with the only positive root being \cite{Zucker2008}
\begin{equation} \label{eq:cubic_root}
    \lambda = \underbrace{\sqrt{N_s^{\ast}}\left[\cos\left(\frac{1}{3}\arccos\left(-\frac{N_s-1}{N_s^{\ast}}\right)\right)-\frac{1}{2}\right]}_{\Theta(N_s)}
\end{equation} 
Further, for $\Theta(N_s)$ defined in \eqref{eq:cubic_root}, observe that $\lambda>\Theta(N_s) \Leftrightarrow f''(\lambda) > 0$ implying that $f(\lambda)$ is convex. Hence, the inequalities \eqref{eq:target_constraint} become
\begin{subequations}\label{eq:target_constraint_2}
\begin{align}
    \hat{\mathbb{F}}(\bvec{x}_i(k), \lambda_{ijk}) & \leq h_{ijk} \label{eq:target_reform_2}\\ 
    \sum_{j=1}^{N_{Tik}} f(\lambda_{ijk}) & \leq \alpha \label{eq:target_lambda_2}\\
    \lambda_{ijk} & \geq \Theta(N_s) \label{eq:lambda_restrict_2}
\end{align}
\end{subequations}
with \eqref{eq:target_reform_2} and \eqref{eq:lambda_restrict_2} iterated over the index $j \in \Nt{1}{N_{Tik}}$. 

In practice, it may be simpler to substitute $\Theta(N_s)$ with $3^{-1/2}$ as $3^{-1/2} \geq \Theta(N_s)$ for all values of $N_s$. Here, $\lambda \leq 3^{-1/2} \Leftrightarrow f(\lambda) \geq 0.75$. In most cases, probabilistic violation thresholds will not take values this large, and outcomes will not be affected as a result.

Formally, the final reformulation is written as \eqref{eq:target_constraint_3}.
\setcounter{equation}{28}

\begin{lem}
The constraint reformulation \eqref{eq:target_constraint_3} will always be convex in $\vec{U}_i$ and $\lambda_{ijk}$. 
\end{lem}

\begin{IEEEproof}
Here, \eqref{eq:target_reform_3} is affine and hence convex, in both $\vec{U}_i$ and $\lambda_{ijk}$. Further, $f(\lambda)$ is convex by the restriction \eqref{eq:lambda_restrict_3}. Since, \eqref{eq:target_lambda_3} is the sum of convex functions, it too is convex. Finally, the control authority $\U$ is a closed and convex set. Therefore, the chance constraint reformulation \eqref{eq:target_constraint_3} will always be convex.
\end{IEEEproof}

\begin{lem} \label{lem:target_satisfy}
For the controllers $\vec{U}_1, \dots, \vec{U}_{N_v}$, if there exists risk allocation variables $\lambda_{ijk}$ satisfying \eqref{eq:target_constraint_3} for constraints in the form of \eqref{eq:constraint_t}, then $\vec{U}_1, \dots, \vec{U}_{N_v}$ satisfies \eqref{eq:prob_1_opt_constraints} almost surely.
\end{lem}

\begin{IEEEproof}
Satisfaction of \eqref{eq:target_reform_3} implies \eqref{eq:enforce_target} holds. Theorem \ref{thm:out_of_sample} upper bounds \eqref{eq:enforce_target}. Boole's inequality and De Morgan's law guarantee that if \eqref{eq:target_lambda} holds then \eqref{eq:prob_1_opt_constraints} is satisfied.
\end{IEEEproof}

We take a moment to discuss Assumption \ref{assm:n_samples}. From \eqref{eq:f_lim}, we see that \eqref{eq:target_lambda_3} is lower bounded by 
\begin{equation}
    \frac{N_{Tik}}{ N_s^{\ast}} \leq {\textstyle \sum_{k=1}^{N}\sum_{i=1}^{N_{Tik}} }  f(\lambda_{ijk}) 
\end{equation}
In theory, this means the number of samples need to be 
\begin{equation}
    N_s \geq \frac{ \sum_{k=1}^{N} N_{Tik}}{\alpha}-1
\end{equation}
such that there may exist a solution that satisfies \eqref{eq:target_lambda_3}. However, since \eqref{eq:f_lim} is an asymptotic bound, more samples will be required to allow for finite values of $\lambda_{ijk}$. In practice, the minimum number of samples needed will be dependent on $\alpha$, the volume of the polytopic region, number of hyperplane constraints, and the magnitude of the variance term.

\subsection{2-Norm Based Collision Avoidance Constraints} \label{ssec:collision_reform}

Next, consider the reformulations of the 2-norm constraints \eqref{eq:constraint_o}-\eqref{eq:constraint_r}. Here, we will derive the reformulation of \eqref{eq:constraint_r}, but the reformulation of \eqref{eq:constraint_o} is nearly identical. Without loss of generality define
\begin{subequations}
\begin{align}
    \vec{z} = & \; S \left(A^k \left(\vec{x}_i(0)-\vec{x}_j(0) \right) + \mathcal{C}(k) \left(\vec{U}_i-\vec{U}_j\right) \right)\\
    \bvec{z}^{[i]} = & \; S \mathcal{D}(k) \left(\bvec{W}_i^{[i]}-\bvec{W}_j^{[i]}\right)  
\end{align}
\end{subequations}
to be the non-stochastic and stochastic element of $S(\bvec{x}^{[i]}_i(k) - \bvec{x}^{[i]}_j (k))$, respectively. Then, we can write the norm as,
\begin{equation}
    \|S(\bvec{x}_i(k) - \bvec{x}_j (k))\| = \| \vec{z} + \bvec{z} \|
\end{equation}
We start by observing 
\begin{equation} \label{eq:norm_square}
\begin{split}
    &\pr{ \bigcap_{i=1}^{N_v-1} \bigcap_{j=i+1}^{N_v}  \bigcap_{k=1}^{N} \| \vec{z} + \bvec{z} \| \geq r } \\
    & \ = \pr{\bigcap_{i=1}^{N_v-1} \bigcap_{j=i+1}^{N_v}  \bigcap_{k=1}^{N} \| \vec{z} + \bvec{z} \|^2 \geq r^2 }    
\end{split}
\end{equation}
as the norm is non-negative. Thus, we can write the 2-norm constraint as
\begin{equation} \label{eq:square_norm}
    \pr{\bigcap_{i=1}^{N_v-1} \bigcap_{j=i+1}^{N_v}  \bigcap_{k=1}^{N} \| \vec{z} + \bvec{z} \|^2 \geq r^2} \geq 1-\gamma
\end{equation}

By taking the complement and applying Boole's inequality,
\begin{equation}
\begin{split}
    &\pr{\bigcup_{i=1}^{N_v-1} \bigcup_{j=i+1}^{N_v}  \bigcup_{k=1}^{N} \| \vec{z} + \bvec{z} \|^2 \leq r^2 } \\
    & \ \leq \sum_{i=1}^{N_v-1} \sum_{j=i+1}^{N_v}  \sum_{k=1}^{N} \pr{ \| \vec{z} + \bvec{z} \|^2 \leq r^2 }    
\end{split}
\end{equation}
Using the approach in \cite{ono2008iterative}, we introduce risk variables $\omega_{ijk}$ to allocate risk to each of the individual probabilities
\begin{subequations}\label{eq:reform_new_var}
\begin{align}
   \pr{ \| \vec{z} + \bvec{z} \|^2 \leq r^2 } & \leq \omega_{ijk}  \\
  \sum_{i=1}^{N_v-1} \sum_{j=i+1}^{N_v}  \sum_{k=1}^{N} \omega_{ijk} &\leq \gamma \\
  \omega_{ijk} & \geq 0 
\end{align}
\end{subequations}

In a similar fashion to Section \ref{ssec:target_reform}, we add an additional constraint based on the sample mean and sample tandard deviation of $\| \vec{z} + \bvec{z} \|^2$ to \eqref{eq:reform_new_var} such that the constraint becomes
\begin{subequations} \label{eq:norm_reform_p1}
\begin{align}
    \pr{\| \vec{z} + \bvec{z} \|^2 \leq r^2} & \leq \omega_{ijk} \label{eq:norm_reform_p1_1} \\
    \hex{\| \vec{z} + \bvec{z} \|^2} - \lambda_{ijk} \hstd{\| \vec{z} + \bvec{z} \|^2} & \geq r^2 \label{eq:norm_reform_p1_2} \\
    \sum_{i=1}^{N_v-1} \sum_{j=i+1}^{N_v}  \sum_{k=1}^{N} \omega_{ijk} & \leq \gamma \\
    \omega_{ijk} & \geq  0 \\
    \lambda_{ijk} & \geq 0
\end{align}
\end{subequations}
For brevity, we denote
\begin{equation}
    \hat{\mathbb{G}}(\bvec{x}_i(k), \bvec{x}_j(k),  \lambda_{ijk}) = \hex{\| \vec{z} + \bvec{z} \|^2} - \lambda_{ijk} \hstd{\| \vec{z} + \bvec{z} \|^2}
\end{equation}
Enforcement of \eqref{eq:norm_reform_p1_2} guarantees that
\begin{equation} \label{eq:norm_bound}
\begin{split}
    & \pr{\| \vec{z} + \bvec{z} \|^2 \leq r^2}   \\
    & \ \leq \pr{\| \vec{z} + \bvec{z} \|^2 \leq \hat{\mathbb{G}}(\bvec{x}_i(k), \bvec{x}_j(k),  \lambda_{ijk}) } 
\end{split}
\end{equation}
Then, by Assumptions \ref{assm:samples}-\ref{assm:samples_not_equal} and Theorem \ref{thm:out_of_sample}, we can substitute
\begin{equation}
    \omega_{ijk} = f(\lambda_{ijk})
\end{equation}
and determine the value for $\lambda_{ijk}$ in terms of $\omega_{ijk}$,
\begin{equation}
    \lambda_{ijk} = \frac{\sqrt{N_s^{\ast}\left(1-\omega_{ijk}\right)}}{\sqrt{N_s \omega_{ijk}}-\sqrt{1-\omega_{ijk}}}
\end{equation}
so long as $\lambda \geq 0$. Here, $\omega_{ijk} > 0 \Leftrightarrow \lambda_{ijk} > 0$. Then, we can write \eqref{eq:norm_reform_p1} as
\begin{subequations} \label{eq:norm_reform_p2} 
\begin{align}
    \pr{\| \vec{z} + \bvec{z} \|^2 \leq r^2} & \leq \omega_{ijk} \label{eq:norm_reform_p2_1}\\
    \hat{\mathbb{G}}\left(\bvec{x}_i(k), \bvec{x}_j(k), \frac{\sqrt{N_s^{\ast}(1-\omega_{ijk})}}{\sqrt{N_s \omega_{ijk}}-\sqrt{1-\omega_{ijk}}}\right) & \geq r^2  \label{eq:norm_reform_p2_2}\\
    \sum_{i=1}^{N_v-1} \sum_{j=i+1}^{N_v}  \sum_{k=1}^{N} \omega_{ijk} & \leq \gamma \\
    \omega_{ijk} & \geq 0
\end{align}
\end{subequations}
Since Theorem \ref{thm:out_of_sample} guarantees that satisfaction of \eqref{eq:norm_reform_p2_2} also satisfies \eqref{eq:norm_reform_p2_1} for any value $\omega_{ijk}>0$, \eqref{eq:norm_reform_p2_1} is redundant and can be removed. The constraint is then
\begin{subequations} \label{eq:norm_reform_p2.5}
\begin{align}
    \hat{\mathbb{G}}\left(\bvec{x}_i(k), \bvec{x}_j(k), \frac{\sqrt{N_s^{\ast}(1-\omega_{ijk})}}{\sqrt{N_s \omega_{ijk}}-\sqrt{1-\omega_{ijk}}}\right)  & \geq r^2  \label{eq:norm_reform_p2.5_2}\\
    \sum_{i=1}^{N_v-1} \sum_{j=i+1}^{N_v}  \sum_{k=1}^{N} \omega_{ijk} & \leq \gamma \\
    \omega_{ijk} & \geq 0
\end{align}
\end{subequations}

\begin{defn}[Difference of Convex Function] \label{defn:dc}
A difference of convex function has the form
\begin{equation}\label{eq:dc_def}
   f(\vec{x})-g(\vec{x}) 
\end{equation}
in which $f, g: \R^n \rightarrow \R$ are convex functions for $\vec{x} \in \R^n$.
\end{defn}

Next, we show that $\hat{\mathbb{G}} (\bvec{x}_i(k), \bvec{x}_j(k), \lambda_{ijk} )$ has a difference of convex and closed form in $\vec{U}_i-\vec{U}_j$. Observe that the sample mean vector and sample variance-covariance matrix of $\bvec{z}$ are 
\begin{subequations}
\begin{align}
    \hex{\bvec{z}} = & \; \frac{1}{N_s} \sum_{[i]=1}^{N_s} \bvec{z}^{[i]} \\
    \hvar{\bvec{z}} 
    = & \; \frac{1}{N_s}  \sum_{[i]=1}^{N_s} \left(\bvec{z}^{[i]} \!-\!\hex{\bvec{z}}\right)\left(\bvec{z}^{[i]} \!-\!\hex{\bvec{z}}\right)^{\top}\! \\
\end{align}
the sample mean and sample variance of $\bvec{z}^{\top}\bvec{z}$ are 
\begin{align}
    \hex{\bvec{z}^{\top}\bvec{z}} = & \; \frac{1}{N_s} \sum_{[i]=1}^{N_s} \bvec{z}^{[i] \top}  \bvec{z}^{[i]} \\
     \hvar{\bvec{z}^{\top}\bvec{z}} 
    = & \; \frac{1}{N_s} \sum_{[i]=1}^{N_s} \left(\bvec{z}^{[i] \top} \bvec{z}^{[i]} \!-\! \hex{\bvec{z}^{\top}\bvec{z}} \right)^2  
\end{align}
and the sample covariance vector between $ \bvec{z}$ and $\bvec{z}^{\top}\bvec{z}$ is
\begin{align}
    \hcov{\bvec{z}}{\bvec{z}^{\top}\bvec{z}} 
    = & \; \frac{1}{N_s} \sum_{[i]=1}^{N_s} \left(\bvec{z}^{[i]} \!-\!\hex{\bvec{z}}\right) \left(\bvec{z}^{[i] \top}\! \bvec{z}^{[i]} \!-\! \hex{\bvec{z}^{\top}\!\bvec{z}} \right)
\end{align}
\end{subequations}
Then the sample mean for the 2-norm constraint is
\begin{subequations} \label{eq:sample_norm_mean}
\begin{align}
    \hex{\| \vec{z} + \bvec{z} \|^2}
    & = \frac{1}{N_s} \sum_{[i]=1}^{N_s} \| \vec{z} + \bvec{z}^{[i]} \|^2 \\
    & = \left\| 
        \begin{bmatrix} I_q & \hex{\bvec{z}} \\ \hex{\bvec{z}}^{\top} & \hex{\bvec{z}^{\top}\bvec{z}} \end{bmatrix}^{\frac{1}{2}}
        \begin{bmatrix} \vec{z} \\ 1 \end{bmatrix}
    \right\|^2 \label{eq:mean_norm}
\end{align}
\end{subequations}
and the sample standard deviation for the 2-norm constraint is
\begin{subequations}\label{eq:sample_norm_var}
\begin{align}
    & \hstd{\| \vec{z} + \bvec{z} \|^2} \\
    & \ = \sqrt{\frac{1}{N_s} \sum_{[i]=1}^{N_s} \left(\| \vec{z} + \bvec{z}^{[i]} \|^2 \!-\! \hex{\| \vec{z} + \bvec{z} \|^2} \right)^2} \\
    & \ = \left\| 
        \begin{bmatrix} 4 \hvar{\bvec{z}} & 2\hcov{\bvec{z}}{\bvec{z}^{\top}\bvec{z}} \\ 2\hcov{\bvec{z}}{\bvec{z}^{\top}\bvec{z}}^{\top} & \hvar{\bvec{z}^{\top}\bvec{z}} \end{bmatrix}^{\frac{1}{2}}
        \begin{bmatrix} \vec{z} \\ 1 \end{bmatrix}
    \right\|  \label{eq:var_norm}
\end{align}
\end{subequations}
Finally, we can write $\hat{\mathbb{G}}\left(\bvec{x}_i(k), \bvec{x}_j(k), \frac{\sqrt{N_s^{\ast}(1-\omega_{ijk})}}{\sqrt{N_s \omega_{ijk}}-\sqrt{1-\omega_{ijk}}}\right)$ as \eqref{eq:long_norm}.
\begin{figure*}
\begin{equation} \label{eq:long_norm}
\begin{split}
    \hat{\mathbb{G}}&\left(\bvec{x}_i(k) , \bvec{x}_j(k), \frac{\sqrt{N_s^{\ast}(1-\omega_{ijk})}}{\sqrt{N_s \omega_{ijk}}-\sqrt{1-\omega_{ijk}}}\right) \\
    = & \underbrace{\left\| 
        \begin{bmatrix} I_q & \hex{\bvec{z}} \\ \hex{\bvec{z}}^{\top} & \hex{\bvec{z}^{\top}\bvec{z}} \end{bmatrix}^{\frac{1}{2}}
        \begin{bmatrix} \vec{z} \\ 1 \end{bmatrix}
    \right\|^2}_{\hex{\| \vec{z} + \bvec{z} \|^2}}  - \underbrace{\frac{\sqrt{N_s^{\ast}(1-\omega_{ijk})}}{\sqrt{N_s \omega_{ijk}}-\sqrt{1-\omega_{ijk}}}}_{\lambda_{ijk}} \underbrace{\left\| 
        \begin{bmatrix} 4 \hvar{\bvec{z}} & 2\hcov{\bvec{z}}{\bvec{z}^{\top}\bvec{z}} \\ 2\hcov{\bvec{z}}{\bvec{z}^{\top}\bvec{z}}^{\top} & \hvar{\bvec{z}^{\top}\bvec{z}} \end{bmatrix}^{\frac{1}{2}}
        \begin{bmatrix} \vec{z} \\ 1 \end{bmatrix}
    \right\|}_{\hstd{\| \vec{z} + \bvec{z} \|^2}} 
\end{split}
\end{equation}
\setcounter{equation}{52}
\hrulefill
\begin{equation}
\begin{split}
    & \frac{\sqrt{N_s^{\ast}(1-\omega_{ijk})}}{\sqrt{N_s \omega_{ijk}}-\sqrt{1-\omega_{ijk}}} \left\|  
        \begin{bmatrix} 
            4  \hvar{\bvec{z}} & 2\hcov{\bvec{z}}{\bvec{z}^{\top}\bvec{z}}  \\
            2 \hcov{\bvec{z}}{\bvec{z}^{\top}\bvec{z}}^{\top} & \hvar{\bvec{z}^{\top} \bvec{z}}
        \end{bmatrix}^{\frac{1}{2}}
        \begin{bmatrix} 
            \vec{z}\\
            1
        \end{bmatrix}
    \right\|  \\
    & \ -
   \underbrace{\left(\left\|
        \begin{bmatrix} 
            I_q &  \hex{\bvec{z}}  \\
             \hex{\bvec{z}}^{\top} & \hex{\bvec{z}^{\top}\bvec{z}}
        \end{bmatrix}^{\frac{1}{2}}
        \begin{bmatrix} 
            \vec{z}^p\\
            1
        \end{bmatrix}\right\|^2 +  2 \left(\vec{z}^p + \hex{\bvec{z}} \right) S \mathcal{C}(k) \left(
            (\vec{U}_i -\vec{U}_j) - (\vec{U}_i^p
            + \vec{U}_j^p) \right)
        \right)}_{\text{First order approximation of }\hex{\|\vec{z} + \bvec{z} \|^2}\text{ based on previous iteration's solution.}}
     \leq -r^2 
\end{split}\label{eq:norm_reform_p6} 
\end{equation}
\hrulefill
\setcounter{equation}{48}
\end{figure*}
Both $\hex{\| \vec{z} + \bvec{z} \|^2}$ and $\hstd{\| \vec{z} + \bvec{z} \|^2}$ are convex terms containing the controller $\vec{U}_i-\vec{U}_j$. Hence, \eqref{eq:long_norm} is a difference of convex function per Definition \ref{defn:dc}. 

Note that \eqref{eq:long_norm} is biconvex \cite{Gorski2007} from the interaction of $\omega_{ijk}$ and $\vec{U}_i-\vec{U}_j$ in $\lambda_{ijk} \hstd{\| \vec{z} + \bvec{z} \|^2}$. For known risk allocation values $\tilde{\omega}_{ijk}$, the constraint \eqref{eq:long_norm} becomes
\begin{equation} \label{eq:norm_reform_p3}
        \hat{\mathbb{G}}\left(\bvec{x}_i(k), \bvec{x}_j(k), \frac{\sqrt{N_s^{\ast}(1-\tilde{\omega}_{ijk})}}{\sqrt{N_s \tilde{\omega}_{ijk}}-\sqrt{1-\tilde{\omega}_{ijk}}}\right) \geq r^2
\end{equation}

\begin{lem} \label{lem:norm_dc}
The constraint \eqref{eq:norm_reform_p3} is always a difference-of-convex function constraint in $\vec{U}_i$ for constraints in the form \eqref{eq:constraint_o} and in $\vec{U}_i - \vec{U}_j$ for constraints in the form \eqref{eq:constraint_r}.
\end{lem}

\begin{IEEEproof}
Observe that $\vec{z}$ is affine in the control input. Then both \eqref{eq:mean_norm} and \eqref{eq:var_norm} are quadratic functions about a positive semi-definite matrix. Hence, both terms are convex. Then \eqref{eq:long_norm} is a difference of convex function per Definition \ref{defn:dc}. As $\tilde{\omega}_{ijk}>0$ and is fixed, it follows that \eqref{eq:norm_reform_p3} is always a difference-of-convex functions constraint.
\end{IEEEproof}

\begin{lem} \label{lem:collision_satisfy}
If the controller $\vec{U}_1, \dots, \vec{U}_{N_v}$, satisfies \eqref{eq:norm_reform_p3} for constraints in the form of \eqref{eq:constraint_o}-\eqref{eq:constraint_r}, then  $\vec{U}_1, \dots, \vec{U}_{N_v}$ satisfy \eqref{eq:prob_1_opt_constraints} almost surely.
\end{lem}

\begin{IEEEproof}
Satisfaction of \eqref{eq:norm_reform_p3} implies \eqref{eq:norm_bound} is satisfied for 
\begin{equation}
    \lambda_{ijk} = \frac{\sqrt{N_s^{\ast}(1-\omega_{ijk})}}{\sqrt{N_s \omega_{ijk}}-\sqrt{1-\omega_{ijk}}}
\end{equation}
Assumption \ref{assm:n_samples} guarantees $\lambda_{ijk}$ is sufficiently tight. Then Theorem \ref{thm:out_of_sample} guarantees satisfaction of \eqref{eq:prob_1_opt_constraints} almost surely. 
\end{IEEEproof}

\subsection{Difference-of-Convex Programming}

Combining the results from Sections \ref{ssec:target_reform} and \ref{ssec:collision_reform}, we obtain a new optimization problem. 
\begin{subequations}\label{prob:big_prob_eq_2}
    \begin{align}
        \underset{\substack{\vec{U}_1, \dots, \vec{U}_{N_v}\\ \lambda_{ijk}}}{\mathrm{minimize}} \quad & J\left(
        \bvec{X}_1, \ldots, \bvec{X}_{N_v},  \vec{U}_1, \dots, \vec{U}_{N_v}\right) \label{eq:dc_cost} \\
        \mathrm{subject\ to} \quad  & \vec{U}_1, \dots, \vec{U}_{N_v} \in  \mathcal U^N,  \\
        & \text{Sample mean and standard deviation} \label{eq:dc_moment}\\
        & \text{terms defined by } \eqref{eq:sample_poly_mean_var} \text{ and } \eqref{eq:sample_norm_mean}-\eqref{eq:sample_norm_var} \nonumber\\
        & \text{Constraints  \eqref{eq:target_constraint_3} and \eqref{eq:norm_reform_p3}} \label{prob:second_eq_prob_constraints} 
    \end{align}
\end{subequations}

\begin{reform} \label{prob:second}
    Under Assumptions \ref{assm:samples}-\ref{assm:n_samples}, solve the stochastic optimization problem \eqref{prob:big_prob_eq_2} with probabilistic violation thresholds $\alpha$, $\beta$, and $\gamma$ for open loop controllers $\vec{U}_1, \dots, \vec{U}_{N_v} \in  \mathcal U^N$ and optimization variables $\lambda_{ijk}$.
\end{reform} 

\begin{lem} \label{lem:conserv}
Solutions to Reformulation \ref{prob:second} are conservative solutions to Problem \ref{prob:1}.
\end{lem}
\begin{IEEEproof}
Lemmas \ref{lem:target_satisfy} and \ref{lem:collision_satisfy} guarantee the probabilistic constraints \eqref{eq:constraints} are satisfied almost surely. Chance constraint bounds provided by Theorem \ref{thm:out_of_sample} are asymptotically convergent (in $N_s$) to Cantelli's inequality. As Cantelli's inequality is conservative, so is Theorem \ref{thm:out_of_sample}. The sample mean and standard deviation terms in Reformulation \ref{prob:second} encompass and replace the dynamics used in Problem \ref{prob:1}. The cost and input constraints remain unchanged.
\end{IEEEproof}

Note that \eqref{prob:big_prob_eq_2} is a difference-of-convex functions optimization problem. A difference-of-convex functions optimization has the form
\begin{equation} \label{eq:dc}
\begin{split}
    \underset{x}{\mathrm{minimize}} \quad & f_0(\vec{x})-g_0(\vec{x})  \\
    \mathrm{subject\ to} \quad  & f_i(\vec{x})-g_i(\vec{x}) \leq 0 \quad \text{for } i \in \mathbb{N}
\end{split}   
\end{equation}
in which $f_0, f_i(\cdot): \R^n \rightarrow \R$ and $g_0, g_i(\cdot): \R^n \rightarrow \R$ for $\vec{x} \in \R^n$ are convex. While \eqref{eq:dc_cost}-\eqref{eq:dc_moment} are convex,  \eqref{prob:second_eq_prob_constraints} is difference of convex due to the constraint \eqref{eq:norm_reform_p3}.

We employ the convex-concave procedure \cite{boyd_dc_2016} to solve \eqref{prob:big_prob_eq_2}. By taking a first order approximation of the 2-norms mean, \eqref{eq:sample_norm_mean}, we can solve the difference of convex function optimization problem iteratively as a convex optimization problem. By updating the first order approximation at each iteration, the convex-concave procedure solves to a local optimum. Here, the first order approximation transforms the difference of convex function constraint \eqref{eq:norm_reform_p3} into the convex constraint \eqref{eq:norm_reform_p6} where the superscript $p$ indicated the value from the previous iteration’s solution. The main benefit of solving this problem with the convex-concave procedure is the first order approximation makes the constraint convex while maintaining the probabilistic assurances.
\setcounter{equation}{53}

Feasibility of \eqref{eq:norm_reform_p6} is dependent on the feasibility of the initial conditions $\vec{z}^p$ and $\vec{U}^p_i$. We also add slack variables to accommodate potentially infeasible initial conditions \cite{boyd_dc_2016,horst2000}. When using a difference of convex functions optimization problem, Lemma \ref{lem:collision_satisfy} guarantees a feasible but locally optimal solution.

\section{Results} \label{sec:results}

We demonstrate our method on a multi-satellite rendezvous problem with two different simulated disturbances that impact the relative satellite dynamics. All computations were done on a 1.80GHz i7 processor with 16GB of RAM, using MATLAB, CVX \cite{cvx} and Gurobi \cite{gurobi}. Polytopic construction and plotting was done with MPT3 \cite{MPT3}. All code is available at \url{https://github.com/unm-hscl/shawnpriore-sample-bound-mpc}.

Consider a scenario in which $N_v$ satellites, called the deputies, are stationed in geostationary Earth orbit, and tasked to rendezvous with a refueling spacecraft, called the chief. The satellites are tasked with reaching a new configuration represented by polytopic target sets. Each deputy must avoid other deputies while navigating to their respective target sets as shown in Figure \ref{fig:demo}. The relative planar dynamics of each deputy, with respect to the position of the chief are described by the CWH equations \cite{wiesel1989_spaceflight}
\begin{subequations}
\begin{align}
\ddot x - 3 \omega^2 x - 2 \omega \dot y &= \frac{F_x}{m_c} \label{eq:cwh:a}\\
\ddot y + 2 \omega \dot x & = \frac{F_y}{m_c} \label{eq:cwh:b}\\
\ddot z + \omega^2 z & = \frac{F_z}{m_c}. \label{eq:cwh:c}
\end{align}   
\label{eq:cwh}
\end{subequations}
with input $\vec{u}_i = [ \begin{array}{ccc} F_x & F_y & F_z\end{array}]^\top$, and orbital rate $\omega = \sqrt{\frac{\mu}{R^3_0}}$, gravitational parameter $\mu$, and orbital radius $R_0$. We discretize \eqref{eq:cwh} under impulse thrust assumptions, with sampling time $\Delta$t$=$ 60s, and insert a disturbance process that captures uncertainties in the model specification, so that dynamics of each satellite are described by  
\begin{equation}
    \bvec{x}_i(k+1) = A \bvec{x}_i(k) + B \vec{u}_i(k) + \bvec{w}_i(k)
\end{equation}

\subsection{Disturbance from Gravitational Effects} \label{ssec:ex_j2}

The CWH equations \eqref{eq:cwh} are the result of a Taylor series expansion of the relative dynamics modeled via the 2-body problem \cite{wiesel1989_spaceflight}. While the 2-body problem models the predominant force in satellite motion, it ignores all the lesser forces that effect satellite motion. Forces such as the $J_2$ effect, solar and lunar third body gravity, and drag can greatly affect the trajectory of the satellite over a long enough time horizon. In the 2-body model, these forces must be considered as disturbances despite lacking a standard distribution form \cite{Prado2003, Chihabi2020}. It is disturbances like these that motivate our approach. 

In this demonstration, we use the disturbance term to capture the three largest disturbances to satellites at altitudes equivalent to geostationary orbit, the $J_2$ effect, and solar and lunar third body gravity. We compare the proposed method with the scenario approach in \cite{calafiore2006scenario, Campi2018TAC}, and the particle control approach in \cite{blackmore2010_particle}. These methods are commonly used to address chance constraints with sample data. Note that neither the scenario approach or the particle control approach can accommodate the 2-norm collision avoidance constraint without embedding an {\em arbitrarily chosen} polytopic approximation of the collision avoidance region. To facilitate a fair comparison, we only consider the convex target set chance constraint so that we need not choose a particular polytopic approximation that may bias the results. 

\subsubsection{Experimental Setup}

\begin{table}
    \caption{Orbital Elements Describing The Initial Orbits Of The Deputy in Section \ref{ssec:ex_j2}. Values of The Deputy are Relative To The Chief's Position.}
    \centering
    \begin{tabular}{lcc}
         \hline \hline
                                    & Chief         & Dep 1                 \\ \hline
         Radius (km)                & 42164.14      & 0.012                 \\ 
         Eccentricity               & 0             & 0                     \\ 
         Inclination ($^\circ$)     & 10            & 9 $\times 10^{-6}$    \\
         RAAN ($^\circ$)            & 0             & 0                     \\
         Arg. of perigee ($^\circ$) & 0             & 0                     \\
         True anomaly ($^\circ$)    & 90            & -8.1$\times10^{-6}$   \\
         \hline
    \end{tabular}
    \label{tab:J2orbits}
\end{table}

We presume there is one deputy with an initial condition given in Table \ref{tab:J2orbits}, the admissible control set is $\U_i = [-0.04, 0.04]^3 N\cdot \Delta$t${}^{-1}$, and time horizon $N=5$, corresponding to 5 minutes of operation. The performance objective is based on fuel consumption. 
\begin{equation}
    J(\vec{U}_1) = \vec{U}_1^\top \vec{U}_1
\end{equation}

\begin{figure}
    \centering
    \includegraphics[width=0.8\columnwidth]{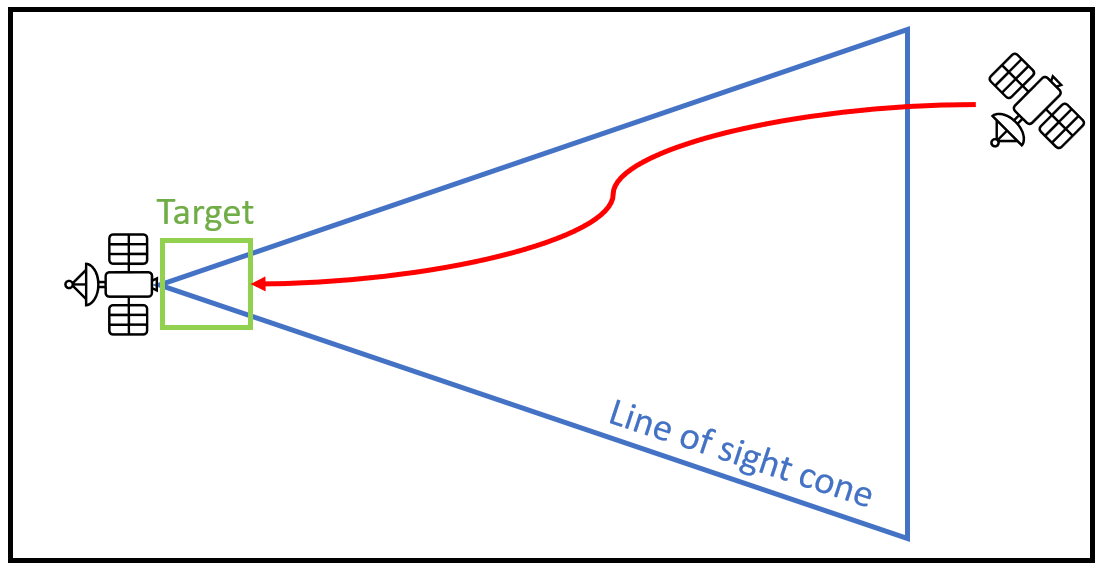}
    \caption{Graphic representation of the problem posed in Section \ref{ssec:ex_j2}. Here, the dynamics of the deputy is perturbed by additive noise representing the $J_2$ gravitational effect, and solar and lunar third body gravity. We attempt to find a control sequence that allows the deputy to rendezvous with the chief while meeting probabilistic time varying target set requirements.}
    \label{fig:problem}
\end{figure}

We choose target sets $\T_1(k)$ that represent a line-of-sight cone for time steps 1 to $N-1$ and a docking position for the terminal time step. The line-of-sight cone for time steps 1 to $N-1$ is defined by
\begin{equation}
    G_k = \begin{bmatrix}
        -1 & 0 & 1 & 0 & 0 & 0 \\
        -1 & 1 & 0 & 0 & 0 & 0 \\
        -1 & 0 & -1 & 0 & 0 & 0 \\
        -1 & -1 & 0 & 0 & 0 & 0 \\
        1 & 0 & 0 & 0 & 0 & 0 
    \end{bmatrix} \; 
    \vec{h}_k = \begin{bmatrix}
        0 \\ 0 \\ 0 \\ 0 \\ 10
    \end{bmatrix}
\end{equation}
The terminal set is defined by 
\begin{equation}
    G_N = I_6 \otimes \begin{bmatrix}
        1 \\ -1
    \end{bmatrix} \; 
    \vec{h}_N = \begin{bmatrix}
        2 & 0 &  \vec{1}_{4}^{\top} & 0.5 \cdot \vec{1}_{6}^{\top}
    \end{bmatrix}^{\top}
\end{equation}
We graphically represent the problem of interest in Figure \ref{fig:problem}. The chance constraint is defined as
\begin{equation}
    \pr{ \bigcap_{k=1}^5 \bvec{x}_1(k) \in \T_{1}(k) } \geq 1-\alpha \label{eq:terminal_j2}
\end{equation}
The violation threshold is chosen to be $\alpha = 0.05$.

We compare all three methods with the same sample set. Because the scenario approach has the largest sample size requirement, we will use its sample size for all three methods. To determine the number of samples needed for the scenario approach, we use the formula
\begin{equation}
    N_s \geq \frac{2}{\alpha}\left( \ln{\frac{1}{\beta}} + N_o \right)
\end{equation}
where $\beta \in (0,1)$ is the confidence bound and $N_o$ is the number of optimization variables \cite{Campi2008}. We set $\beta = 10^{-16}$ and observe that $N_o = 15$, hence, we use $N_s = 2,073$ samples. 

\begin{figure}
    \centering
    \includegraphics[width=\columnwidth]{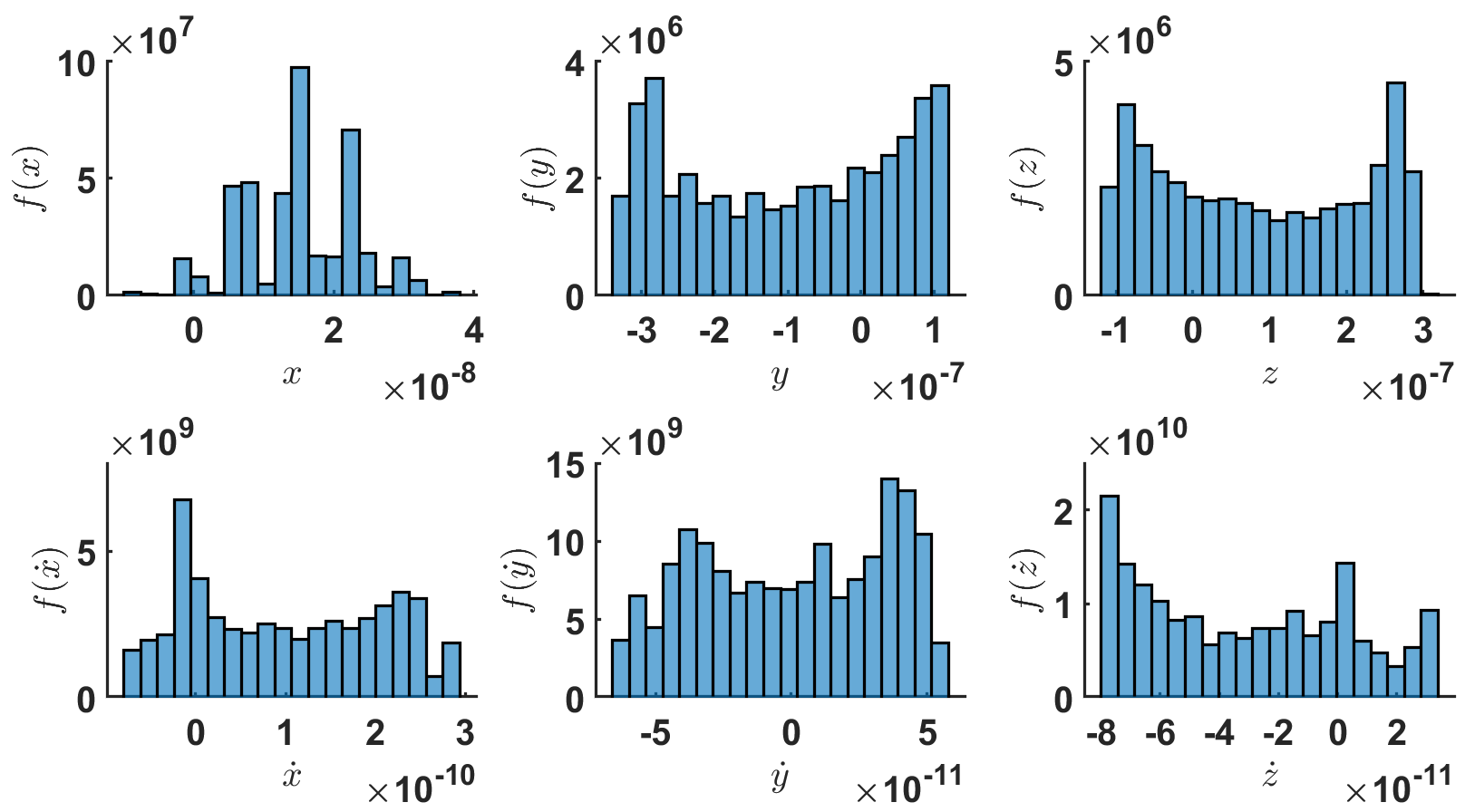}
    \caption{Histograms of $N_s=2,073$ sampled disturbances for each element of the Deputy's disturbance vector at time step 0, $\bvec{w}_1^{[i]}(0)$. Notice they all have highly irregular shapes.}
    \label{fig:dist}
\end{figure}

To generate our disturbance data, we propagate the chief and the deputy in one time step increments via both the 2-body equations and the force model:
\begin{equation}
    \ddot{\vec{x}} = \frac{1}{m} \left( F_{GM} + F_{J2} + F_{Sun} + F_{Moon} \right)
\end{equation}
where $F_{GM}$ is the 2-body force, $F_{J2}$ is the force created by the $J_2$ gravitational harmonic, and $F_{Sun}$ and $F_{Moon}$ are the third body gravity forces from the Sun and Moon, respectively. For the equations of motion, and the position calculations for the Sun and Moon, we follow Sections 3.2 and 3.3 of \cite{Montenbruck2000}. We do not include them here for brevity. We then convert the propagated trajectories into the chief's body fixed local frame and take the difference as the disturbance. This process was completed for $N_s=2,073$ sample disturbances with sun and moon positions calculated with a random and uniformly distributed time between midnight on November 13\textsuperscript{th}, 2022 and January 12\textsuperscript{th}, 2023. Figure \ref{fig:dist} plots a histogram for each element of the Deputy's disturbance vector at time step 0, $\bvec{w}_1^{[i]}(0)$. As we can see, each of the histograms display very different and non-Gaussian disturbances. Each of these distributions would be challenging to characterize with standard distributions, leading to poor results from modeling-based approaches. 

\subsubsection{Experimental Results}

\begin{figure}
    \centering
    \includegraphics[width=\columnwidth]{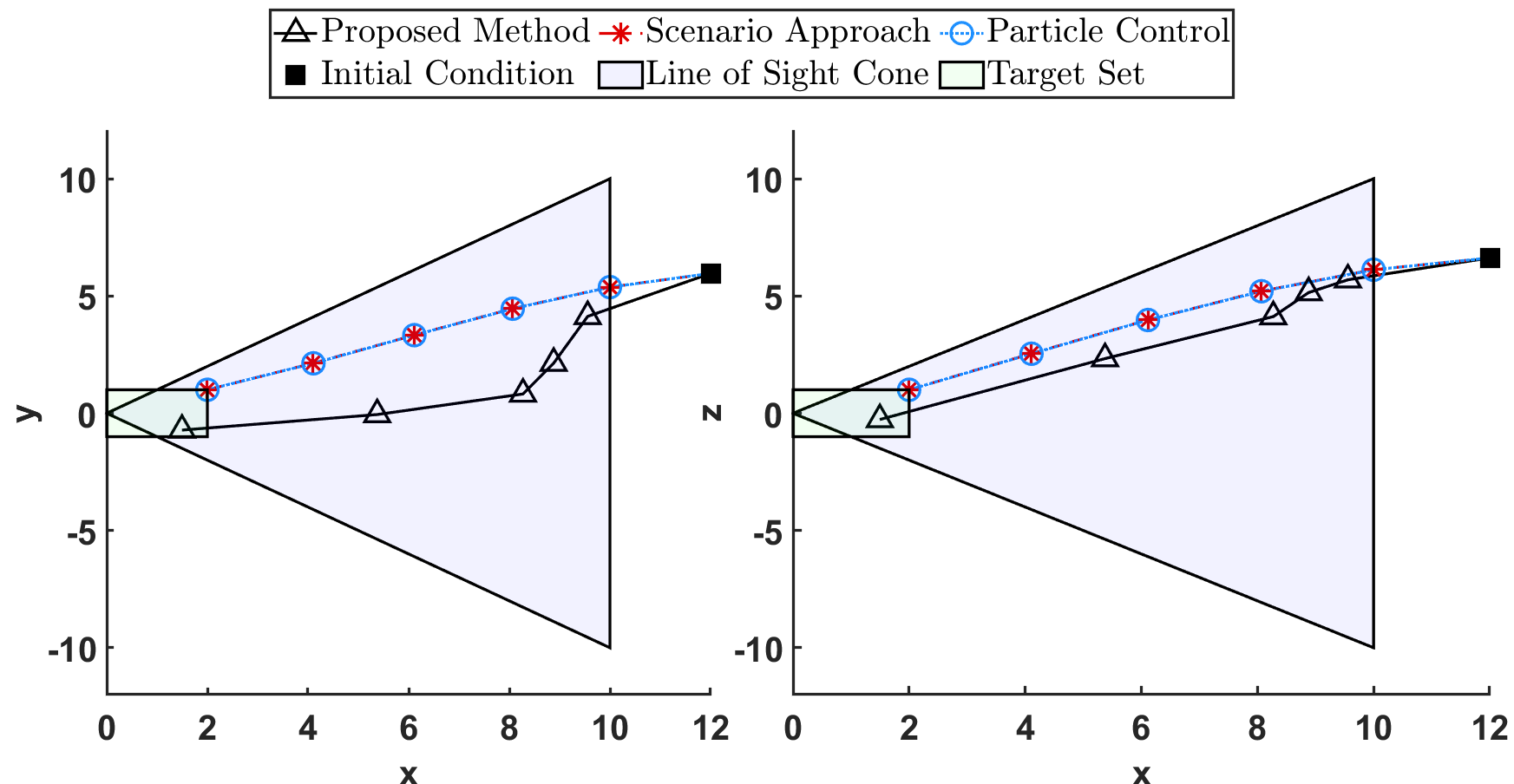}
    \caption{Comparison of mean trajectories between proposed method (solid line) and robust control approach (dashed line) for CWH dynamics. Disturbances were sampled from the difference in the CWH frame between 2-body dynamics and dynamics with 2-body acceleration, $J_2$ gravitational acceleration, and solar and lunar acceleration.}
    \label{fig:hf}
\end{figure}

Figure \ref{fig:hf} shows the resulting trajectories of the three methods. We see that while the trajectories of the scenario approach and particle control are similar, the proposed method resulted in a trajectory that is less smooth. The lack of smoothness is an embodiment of the conservatism present in our approach. Solution statistics and empirical chance constraint satisfaction can be found in Table \ref{tab:cwh_stats}. We see the solution cost was larger for the proposed method, as expected. To assess constraint satisfaction, we generated $10^4$ additional disturbances and empirically tested whether the target set constraint was satisfied. We expect the proposed method to always empirically satisfy the chance constraint, given Lemma \ref{lem:target_satisfy}. However, this guarantee cannot be made for either the scenario approach or the particle control approach. While in this instance, the scenario approach did satisfy the chance constraint empirically, the particle control approach did not.

We point out in Table \ref{tab:cwh_stats} the large difference in computation time between the three methods. The time to solve the solution with the proposed method is \textit{two orders of magnitude faster than both the scenario approach and the particle control approach}. Here, the computational benefits and almost surely guarantees of chance constraint satisfaction present a strong case to use this method in instances where chance constraint satisfaction and speed are important.  

\begin{table*}
    \caption{Solution Cost, Computation Time, and Empirical Constraint Satisfaction for CWH Dynamics with Simulated $J_2$, Sun, and Moon Acceleration Disturbance and probabilistic violation threshold $\alpha = 0.05$. Constraint Satisfaction is the Ratio of $10^4$ Additional Simulated Disturbance Samples that Satisfied the Constraint.}   
    \centering
    \begin{tabular}{lccc}
         \hline \hline
         Metric                 & Proposed Method  & Scenario Approach \cite{calafiore2006scenario, Campi2018TAC} & Particle Control \cite{blackmore2010_particle} \\ \hline
         Solve Time (sec)                           & 0.3472    & 15.7706   &  64.8701  \\ 
         Solution Cost ($N^2 \times 10^{-3}$)       & 6.1954    & 1.5832    &  1.5824   \\ 
         Target Sets \eqref{eq:terminal_j2}           & 1.0000    & 1.0000    &  0.2061   \\ \hline
    \end{tabular}
    \label{tab:cwh_stats}
\end{table*}

\subsection{Gaussian Disturbance}

We compare our method against the method in \cite{paulson2017stochastic}, the asymptotic and analytic counterpart of the method proposed in this work based on Cantelli's inequality. This approach is effective for and has been demonstrated on systems which have target constraints and can be solved via convex optimization. We extend this method to accommodate 2-norm based collision constraints (as in Section \ref{ssec:collision_reform}) for the purpose of comparison with our own approach.  

\begin{thm}[Cantelli's inequality \cite{Boucheron2013}]
Let $\boldsymbol{x}$ be a real valued random variable with finite expectation $\ex{\boldsymbol{x}}$ and finite, non-zero standard deviation $\std{\boldsymbol{x}}$. Then, for any $\lambda > 0$, 
\end{thm}
\begin{equation} \label{eq:cantelli}
    \pr{\boldsymbol{x} - \ex{\boldsymbol{x}}  \geq  \lambda \std{\boldsymbol{x}}} \leq \frac{1}{\lambda^2+1}
\end{equation}

As mentioned in the proof of Lemma \ref{lem:conserv}, Theorem \ref{thm:out_of_sample} is asymptotically convergent in $N_s$ to the Cantelli's inequality \cite{Boucheron2013} inequality per the central limit theorem \cite{casella2002}. In this demonstration, we show that Theorem \ref{thm:out_of_sample} does not add significant conservatism in comparison to Cantelli's inequality despite using sample statistics over analytic expressions of moments. This can be particularly useful as computing sample statistics requires little computational overhead and no knowledge of the underlying disturbance. 

\subsubsection{Experimental Setup}

We presume there are three deputies with the initial conditions are given by Table \ref{tab:Gaussorbits}, the admissible control set is $\U_i = [-4, 4]^3 N\cdot \Delta$t${}^{-1}$, and time horizon $N=5$, corresponding to 5 minutes of operation. The performance objective is based on fuel consumption. 
\begin{equation}
    J(\vec{U}_1, \dots, \vec{U}_3) = \sum^3_{i=1} \vec{U}_i^\top \vec{U}_i
\end{equation}

We presume the disturbance is a zero-mean Gaussian distribution,
\begin{equation} \label{eq:gauss_param}
    \boldsymbol{W}_i \sim \mathcal{N}(\vec{0}, I_5 \otimes \mathrm{blkdiag}(10^{-5} \cdot I_3, 10^{-8} \cdot I_3 ))
\end{equation}
Here, we have selected the sample size for the proposed method to be $N_s = 5,000$. For comparison, we use the expectation and covariance matrix parameters\eqref{eq:gauss_param} to compute a solution with the method of \cite{paulson2017stochastic}. 

The terminal sets $\T_i(N)$ are $5\times 5 \times 5$m boxes centered around desired terminal locations in $x,y$ coordinates approximately 11m away from the origin, with velocity bounded in all directions by $[-0.25, 0.25]$m/s. For collision avoidance, we presume that the deputies must remain at least $r=10$m away from each other and the chief, hence $S = \begin{bmatrix} I_{3} & 0_{3} \end{bmatrix}$ to extract the positions. Violation thresholds for terminal sets and collision avoidance are $\alpha = \beta = \gamma = 0.05$, respectively. The chance constraints are defined as
\begin{align}
    \pr{ \bigcap_{i=1}^3 \bvec{x}_i(N) \in \T_{i}(N) } &\geq 1-\alpha \label{eq:terminal}\\
    \pr{ \bigcap_{k=1}^{5} \bigcap_{i=1}^{3} \left\| S \bvec{x}_i(k)\right\| \geq r } &\geq 1-\beta  \label{eq:avoidance_chief} \\
    \pr{ \bigcap_{k=1}^{5} \bigcap_{i,j=1}^{3} \left\| S \left(\bvec{x}_i(k) -  \bvec{x}_j(k)\right)\right\| \geq r } &\geq 1-\gamma  \label{eq:avoidance_dep}
\end{align}
Here $S = \begin{bmatrix} I_3 & 0_3 \end{bmatrix}$.

As has been established \cite{ono2008iterative, Gorski2007}, biconvexity associated with having both risk allocation and control variables can be addressed in an iterative fashion, by alternately solving for the risk allocation variables, then for the control. However, for our demonstration, to isolate the impact of Theorem \ref{thm:out_of_sample}, we presume a fixed risk allocation. We uniformly allocate risk such that 
\begin{align}
    \pr{  \left\| S \boldsymbol x_i(k) \right\| \geq r } &\geq 1-\hat{\beta} & \forall  \; i,k \label{eq:avoidance_RV_each} \\
    \pr{ \left\| S \left(\boldsymbol x_i(k) -  \boldsymbol x_j(k)\right)\right\| \geq r } &\geq 1-\hat{\gamma} & \forall \; i,j,k \label{eq:avoidance_AV_each}
\end{align}
where $\hat{\gamma} = \frac{\gamma}{15 \text{ constraints}} = \frac{.05}{15} = 0.00\overline{3}$ and similarly $\hat{\beta} = 0.00\overline{3}$. These values remain constant throughout the iterative solution finding process.

We define the solution convergence thresholds for the convex-concave procedure as both the difference of sequential performance objectives as less than $10^{-6}$ and the sum of slack variables as less than $10^{-8}$. Difference of convex programs were limited to 100 iterations. The first order approximations of the reverse convex constraints were instantiated with no system input.

\begin{table}[b]
    \caption{Orbital Elements Describing The Initial Orbits Of Each Satellite. Values For The Deputies Are Relative To The Chief's Position.}
    \centering
    \begin{tabular}{lcccc}
         \hline \hline
                                & Chief & Dep 1 & Dep 2 & Dep 3   \\ \hline
         Radius (km)   & 42164.14 & 0.080 & 0.085 & 0.087         \\ 
         Eccentricity           & 0 & 0 & 0 & 0               \\ 
         Inclination ($^\circ$) & 10 & -2$\times 10^{-5}$ & 0 & $10^-5$       \\
         RAAN ($^\circ$)        & 0 & 0 & 0 & 0 \\
         Arg. of perigee ($^\circ$) & 0 & 0 & 0 & 0 \\
         True anomaly ($^\circ$) & 90 &   1.8$\times10^{-5}$ & -9$\times10^{-6}$ & 9$\times10^{-6}$\\
         \hline
    \end{tabular}
    \label{tab:Gaussorbits}
\end{table}

\subsubsection{Experimental Results}

\begin{figure}
    \centering
    \includegraphics[width=0.95\columnwidth]{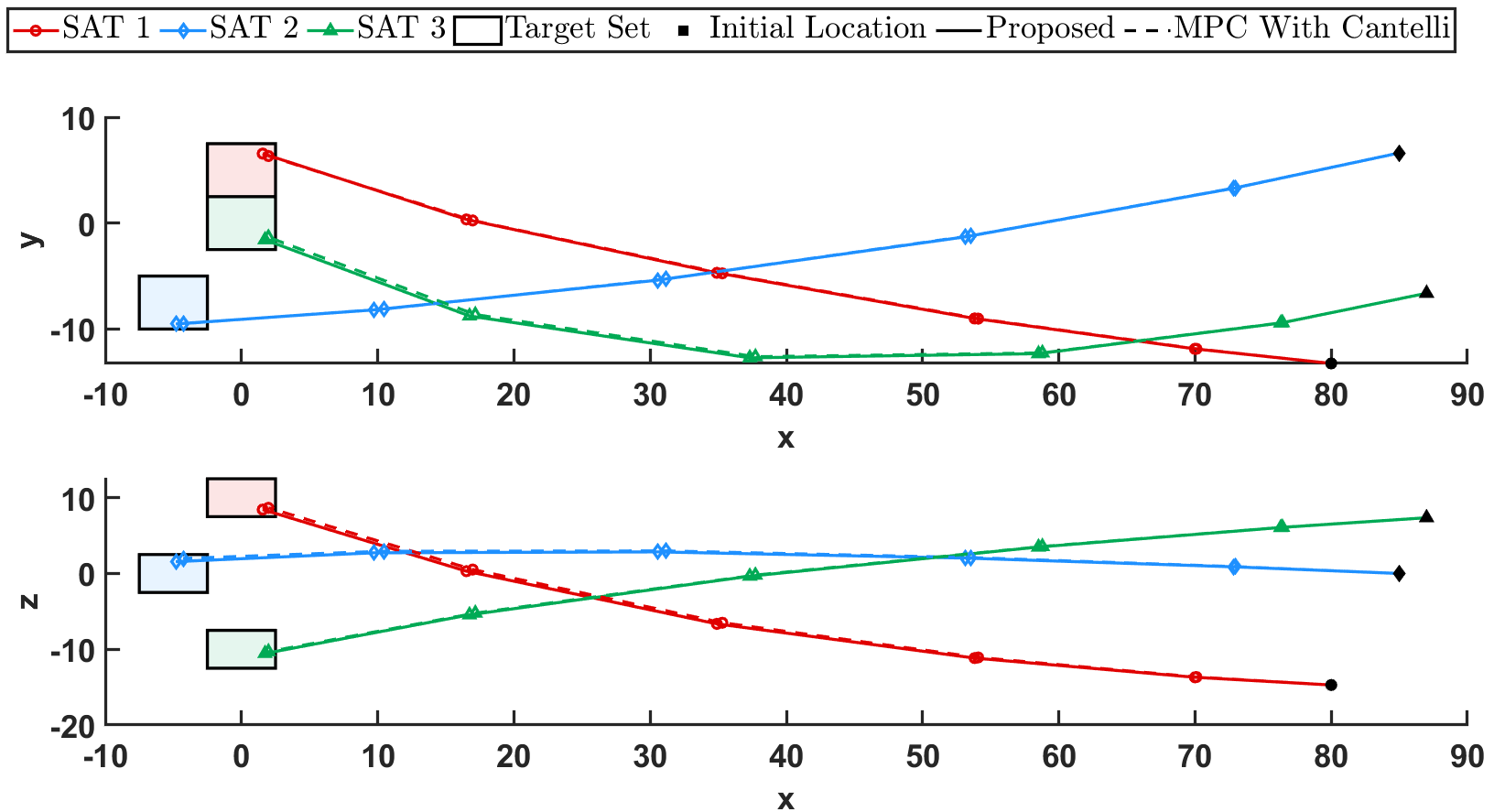}
    \caption{Comparison of mean trajectories between proposed method (solid line) and MPC approach using Cantelli's inequality \cite{paulson2017stochastic} (dashed line) for CWH dynamics with Gaussian disturbance. Notice the trajectories are nearly identical. }
    \label{fig:gauss}
\end{figure}

Figure \ref{fig:gauss} shows the resulting trajectories of the two methods and Table \ref{tab:gauss} compares the time to compute a solution, solution cost, and empirical chance constraint satisfaction with $10^4$ additional samples disturbances. The two methods preformed near identically. The only notable difference is that the proposed method resulted in a  approximate 2\% increase in solution cost. This is a small increase if we consider the proposed method does not require full knowledge of the underlying distribution. Here, we have shown that the proposed method results in only a small deviation for a finite sample size in comparison to that of its asymptotic and analytic counterpart as in \cite{paulson2017stochastic}.

\begin{table}
    \caption{Solution Cost, Computation Time, and Empirical Constraint Satisfaction for CWH Dynamics with Gaussian Disturbance and probabilistic violation thresholds $\alpha = \beta = \gamma = 0.05$. Constraint Satisfaction is Measured as a Ratio of $10^4$ Monte Carlo Disturbance Samples that Satisfy the Constraint.}   
    \centering
    \begin{tabular}{lcc}
        \hline \hline
        Metric                 & Proposed    & MPC with Cantelli's Ineq. \cite{paulson2017stochastic} \\ \hline
        Solve Time (sec)       & 6.9472             & 6.9833  \\ 
        Iterations             & 8                  & 8   \\ 
        Solution Cost ($N^2$)  & 0.2020             & 0.1971   \\
        Terminal Set \eqref{eq:terminal} & 1.0000 & 1.0000 \\
        Avoid the Chief \eqref{eq:avoidance_chief} & 1.0000 & 1.0000 \\
        Avoid Each Other \eqref{eq:avoidance_dep} & 1.0000 & 1.0000 \\\hline
    \end{tabular}
    \label{tab:gauss}
\end{table}

\section{Conclusions and Future Work} \label{sec:conclusion}

We proposed a method based on sample statistics to solve chance constrained stochastic optimal control problems with almost surely guarantees of chance constraint satisfaction. This work focuses on probabilistic requirements for polytopic target sets and 2-norm based collision avoidance constraints in disturbed LTI systems. We derived a concentration inequality that allow us to bound tail probabilities of a random variable being a set number of sample standard deviations away from the sample mean. Our approach relies on this derived concentration inequality to reformulate joint chance constraints into a series of inequalities that can be readily solved as a difference of convex functions optimization problem. We demonstrated our method on two multi-satellite rendezvous scenarios. The first scenario was modeled with disturbance data generated from the $J_2$ gravitational harmonic, and third body gravity from the Sun and Moon, and compared against the scenario approach and particle control. The second scenario was modeled with a zero-mean Gaussian disturbance and compared against a model predictive control approach using Cantelli's inequality, the analytic analogue of Theorem \ref{thm:out_of_sample}. In the two examples, we showed that this approach is amenable to probabilistic guarantees, is efficient to compute, and may be a effective alternative to moment based model predictive control methods.

In future work, we are interested in exploring sample-based concentration inequalities that result in less conservative bounds. We are currently exploring an extension of the methodology presented for unimodal disturbances.

\appendix

\subsection{Proof of Theorem 1} \label{appx:out-sample}

To prove Theorem \ref{thm:out_of_sample}, we first need to state and prove the following Lemma. 

\begin{lem} \label{thm:in_sample}
Let $\boldsymbol{x}^{[1]},\dots, \boldsymbol{x}^{[N_s]}$ be identically distributed samples drawn independently where $N_s \geq2$. Let
\begin{subequations}
\begin{align}
    \hex{\boldsymbol{x}} = & \; \frac{1}{N_s} \sum_{i=1}^{N_s} \boldsymbol{x}^{[i]} \\
    \hstd{\boldsymbol{x}} =  & \; \sqrt{\frac{1}{N_s} \sum_{i=1}^{N_s} (\boldsymbol{x}^{[i]} - \hex{\boldsymbol{x}})^2}
\end{align}
\end{subequations}
be the sample mean and sample standard deviation, respectively, with $\hstd{\boldsymbol{x}}>0$ almost surely. Then for $\lambda>0$
\begin{equation} \label{eq:within_sample_cantelli}
    \pr{\boldsymbol{x}^{[i]}-\hex{\boldsymbol{x}} \geq \lambda \hstd{\boldsymbol{x}}} \leq \frac{1}{\lambda^2 + 1}
\end{equation}
for $i \in \Nt{1}{N_s}$.
\end{lem}

\begin{IEEEproof}
Observe
\begin{subequations}
\begin{align}
    & \pr{\boldsymbol{x}^{[i]} - \hex{\boldsymbol{x}} \geq \lambda \hstd{\boldsymbol{x}}}\\
    & \; =  \pr{\frac{\boldsymbol{x}^{[i]} - \hex{\boldsymbol{x}}}{\hstd{\boldsymbol{x}}} \geq \lambda } \\
    & \; =  \pr{\frac{\boldsymbol{x}^{[i]} \!-\! \hex{\boldsymbol{x}}}{\hstd{\boldsymbol{x}}} - \ex{\frac{\boldsymbol{x}^{[i]} \!-\! \hex{\boldsymbol{x}}}{\hstd{\boldsymbol{x}}}} \geq \lambda \std{\frac{\boldsymbol{x}^{[i]} \!-\! \hex{\boldsymbol{x}}}{\hstd{\boldsymbol{x}}}} } 
\end{align}
\end{subequations}
as 
\begin{subequations}
    \begin{align}
        \ex{\frac{\boldsymbol{x}^{[i]} \!-\! \hex{\boldsymbol{x}}}{\hstd{\boldsymbol{x}}}} & = 0 \\
        \std{\frac{\boldsymbol{x}^{[i]} \!-\! \hex{\boldsymbol{x}}}{\hstd{\boldsymbol{x}}}} & = 1
    \end{align}
\end{subequations}
Then by Cantelli's inequality
\begin{subequations}
\begin{align}
    & \pr{\frac{\boldsymbol{x}^{[i]} \!-\! \hex{\boldsymbol{x}}}{\hstd{\boldsymbol{x}}} - \ex{\frac{\boldsymbol{x}^{[i]} \!-\! \hex{\boldsymbol{x}}}{\hstd{\boldsymbol{x}}}} \geq \lambda \std{\frac{\boldsymbol{x}^{[i]} \!-\! \hex{\boldsymbol{x}}}{\hstd{\boldsymbol{x}}}} } \\
    & \leq \frac{1}{\lambda^2 + 1}
\end{align}
\end{subequations}
for any $\lambda >0$.
\end{IEEEproof}

With Lemma \ref{thm:in_sample} established, we can prove Theorem \ref{thm:out_of_sample}.

\begin{IEEEproof}[Proof of Theorem \ref{thm:out_of_sample}]
Let $\hex{\boldsymbol{x}}^{\ast}$ and $\hstd{\boldsymbol{x}}^{\ast}$ be the sample mean and sample standard deviation calculates with $N_s^{\ast} = N_s + 1$ samples. Note that
\begin{subequations}
\begin{align}
    \boldsymbol{x}^{[N_s^{\ast}]}-\hex{\boldsymbol{x}}^{\ast} & = \; \frac{N_s}{N_s^{\ast}}(\boldsymbol{x}^{[N_s^{\ast}]}-\hex{\boldsymbol{x}}) \\
    N_s^{\ast2} \hvar{\boldsymbol{x}}^{\ast}  & = \; N_s N_s^{\ast} \hvar{\boldsymbol{x}} \!+\!  N_s (\boldsymbol{x}^{[N_s^{\ast}]}-\hex{\boldsymbol{x}})^2 \label{eq:sample_var_conversion}
\end{align}
\end{subequations}
%The derivation of \eqref{eq:sample_var_conversion} is shown in Appendix \ref{appx:deriv_sample_conversion}. 
Then for $\lambda >0$
\begin{subequations}
\begin{align}
    & \pr{\boldsymbol{x}^{[N_s^{\ast}]}-\hex{\boldsymbol{x}} \geq \lambda \hstd{\boldsymbol{x}}} \\
    & \; = \pr{ \sqrt{N_s N_s^{\ast}}(\boldsymbol{x}^{[N_s^{\ast}]}\!-\!\hex{\boldsymbol{x}}) \geq  \lambda \sqrt{N_s N_s^{\ast}} \hstd{\boldsymbol{x}}} \\
    & \; = \P\left((\sqrt{N_s N_s^{\ast}}+\lambda\sqrt{N_s})(\boldsymbol{x}^{[N_s^{\ast}]}-\hex{\boldsymbol{x}}) \right.\\ 
    & \qquad \quad \left. \geq \lambda \sqrt{N_s N_s^{\ast}}\hstd{\boldsymbol{x}} +  \lambda\sqrt{N_s}(\boldsymbol{x}^{[N_s^{\ast}]}-\hex{\boldsymbol{x}})\right) \nonumber  \\
    & \; \leq \P\left((\sqrt{N_s N_s^{\ast}}+\lambda\sqrt{N_s})(\boldsymbol{x}^{[N_s^{\ast}]}-\hex{\boldsymbol{x}})\right. \label{eq:out_of_sample_triangle}\\
    & \qquad \quad \left.\geq \lambda \sqrt{N_s N_s^{\ast}\hvar{\boldsymbol{x}} + N_s (\boldsymbol{x}^{[N_s^{\ast}]}-\hex{\boldsymbol{x}})^2}\right) \nonumber
\end{align}
where \eqref{eq:out_of_sample_triangle} results from the triangle inequality. Then,
\begin{align}    
    & \P\left((\sqrt{N_s N_s^{\ast}}+\lambda\sqrt{N_s})(\boldsymbol{x}^{[N_s^{\ast}]} -\hex{\boldsymbol{x}})\right. \nonumber\\
    & \qquad \quad \left.\geq \lambda \sqrt{N_s N_s^{\ast}\hvar{\boldsymbol{x}} + N_s(\boldsymbol{x}^{[N_s^{\ast}]}-\hex{\boldsymbol{x}})^2}\right) \nonumber\\
    & \; =\P\left((\sqrt{N_s N_s^{\ast}}+\lambda\sqrt{N_s})(\boldsymbol{x}^{[N_s^{\ast}]}-\hex{\boldsymbol{x}})\right. \\
    &\qquad \quad \left.\geq \lambda N_s^{\ast}\hstd{\boldsymbol{x}}^{\ast} \right) \nonumber \\
    & \; = \pr{(\boldsymbol{x}^{[N_s^{\ast}]}-\hex{\boldsymbol{x}}) \geq \frac{\lambda N_s^{\ast}}{ \sqrt{N_s N_s^{\ast}}+\lambda\sqrt{N_s}}\hstd{\boldsymbol{x}}^{\ast} } \\
    & \; = \pr{\frac{N_s}{N_s^{\ast}}(\boldsymbol{x}^{[N_s^{\ast}]}-\hex{\boldsymbol{x}}) \geq \frac{\lambda \sqrt{N_s}}{\sqrt{N_s^{\ast}}+\lambda}\hstd{\boldsymbol{x}}^{\ast} } \\
    & \; = \pr{\boldsymbol{x}^{[N_s^{\ast}]}-\hex{\boldsymbol{x}}^{\ast} \geq \frac{\lambda \sqrt{N_s}}{\sqrt{N_s^{\ast}}+\lambda}\hstd{\boldsymbol{x}}^{\ast} } \\
    & \; = \pr{\boldsymbol{x}^{[N_s^{\ast}]}-\hex{\boldsymbol{x}}^{\ast} \geq \kappa \hstd{\boldsymbol{x}}^{\ast}} 
\end{align}
Where $\kappa$ is a simple substitution. Here, $\lambda >0$ implies $\kappa>0$. So, by Lemma \ref{thm:in_sample},
\begin{align}
    & \pr{\boldsymbol{x}^{[N_s^{\ast}]}-\hex{\boldsymbol{x}}^{\ast} \geq \kappa \hstd{\boldsymbol{x}}^{\ast}} \nonumber \\
    & \; \leq \frac{1}{\kappa^2 + 1} \\
    & \; = \frac{1}{\left(\frac{\lambda \sqrt{N_s}}{\sqrt{N_s^{\ast}}+\lambda}\right)^2 + 1} \label{eq:out_of_sample_result}
\end{align}
\end{subequations}
Simplifying \eqref{eq:out_of_sample_result} leads to \eqref{eq:out_sample_cantelli}.
\end{IEEEproof}

% \subsection{Derivation of (\ref{eq:sample_var_conversion})} \label{appx:deriv_sample_conversion}
% \begin{subequations}
% \begin{align}
%     & s_{n+1}^2 \\
%     & = \; \frac{1}{n+1}\sum_{i=1}^{n+1} (\boldsymbol{x}^{[i]} - m_{n+1})^2 \\
%     & = \; \frac{1}{n+1}\sum_{i=1}^{n+1} \boldsymbol{x}^{[i]2} - m_{n+1}^2 \\
%     & = \; \frac{1}{n+1}\sum_{i=1}^{n+1} \boldsymbol{x}^{[i]2} - \frac{1}{(n+1)^2}(\sum_{i=1}^{n} \boldsymbol{x}^{[i]} + \boldsymbol{x}^{[n+1]})^2 \\
%     & = \; \frac{1}{n+1}\sum_{i=1}^{n+1} \boldsymbol{x}^{[i]2} - \frac{n^2}{(n+1)^2}m_n^2 - \frac{2n}{(n+1)^2} m_n \boldsymbol{x}^{[n+1]} \nonumber\\
%     & \qquad - \frac{1}{(n+1)^2} \boldsymbol{x}^{[n+1]2}\\
%     & = \; \frac{1}{n+1}\sum_{i=1}^{n+1} \boldsymbol{x}^{[i]2} - \frac{n^2+n}{(n+1)^2}m_n^2 - \frac{n+1}{(n+1)^2} \boldsymbol{x}^{[n+1]2} \nonumber \\
%     & \qquad + \frac{n}{(n+1)^2}(\boldsymbol{x}^{[n+1]}-m_n)^2 \\
%     & = \; \frac{1}{n+1}\left(\sum_{i=1}^{n} \boldsymbol{x}^{[i]2} - nm_n^2\right) + \frac{n}{(n+1)^2}(\boldsymbol{x}^{[n+1]}-m_n)^2 \\
%     & = \; \frac{n}{n+1}s_n^2 + \frac{n}{(n+1)^2}(\boldsymbol{x}^{[n+1]}-m_n)^2 
% \end{align}
% \end{subequations}

\bibliographystyle{ieeetr}
\bibliography{main}

\begin{thebibliography}{10}

\bibitem{blackmore2010_particle}
L.~Blackmore, M.~Ono, A.~Bektassov, and B.~C. Williams, ``A probabilistic
  particle-control approximation of chance-constrained stochastic predictive
  control,'' {\em IEEE Trans. on Robotics}, vol.~26, pp.~502--517, June 2010.

\bibitem{Blackmore2006}
L.~Blackmore, ``A probabilistic particle control approach to optimal, robust
  predictive control,'' in {\em Proceedings of the AIAA Guidance, Navigation,
  and Control Conference and Exhibit}, AIAA, 2006.

\bibitem{calafiore2006scenario}
G.~Calafiore and M.~Campi, ``The scenario approach to robust control design,''
  {\em {IEEE} Trans. Autom. Control}, vol.~51, no.~5, pp.~742--753, 2006.

\bibitem{Campi2008}
M.~C. Campi, S.~Garatti, and M.~Prandini, ``The scenario approach for systems
  and control design,'' {\em IFAC Proceedings Volumes}, vol.~41, no.~2,
  pp.~381--389, 2008.
\newblock 17th IFAC World Congress.

\bibitem{campi2011sampling}
M.~Campi and S.~Garatti, ``A sampling-and-discarding approach to
  chance-constrained optimization: Feasibility and optimality,'' {\em J. Optim
  Theory Appl.}, vol.~148, no.~2, pp.~257--280, 2011.

\bibitem{care2014fast}
A.~Car{\`e}, S.~Garatti, and M.~C. Campi, ``Fast--fast algorithm for the
  scenario technique,'' {\em Ops. Res.}, vol.~62, no.~3, pp.~662--671, 2014.

\bibitem{Saha2010}
S.~Saha, E.~\"{O}zkan, F.~Gustafsson, and V.~\v{S}m\'{\i}dl, ``Marginalized
  particle filters for bayesian estimation of gaussian noise parameters,'' in
  {\em 2010 13th International Conference on Information Fusion}, pp.~1--8,
  2010.

\bibitem{Madankan15}
R.~Madankan, P.~Singla, and T.~Singh, ``A robust data assimilation approach in
  the absence of sensor statistical properties,'' in {\em 2015 American Control
  Conference (ACC)}, pp.~5206--5211, 2015.

\bibitem{RAJAMANI09}
M.~R. Rajamani and J.~B. Rawlings, ``Estimation of the disturbance structure
  from data using semidefinite programming and optimal weighting,'' {\em
  Automatica}, vol.~45, no.~1, pp.~142--148, 2009.

\bibitem{Ben-Tal2009}
A.~Ben-Tal, L.~El~Ghaoui, and A.~Nemirovski, {\em Robust Optimization}.
\newblock Princeton Series in Applied Mathematics, Princeton University Press,
  October 2009.

\bibitem{Lam15}
C.-P. Lam, A.~Y. Yang, and S.~S. Sastry, ``An efficient algorithm for
  discrete-time hidden mode stochastic hybrid systems,'' in {\em 2015 European
  Control Conference (ECC)}, pp.~1212--1218, 2015.

\bibitem{Verma10}
R.~Verma and D.~Del~Vecchio, ``Control of hybrid automata with hidden modes:
  Translation to a perfect state information problem,'' in {\em 49th IEEE Conf.
  Dec. \& Control}, pp.~5768--5774, 2010.

\bibitem{casella2002}
G.~Casella and R.~Berger, {\em Statistical Inference}.
\newblock Duxbury advanced series in statistics and decision sciences, Cengage
  Learning, 2002.

\bibitem{ono2008iterative}
M.~Ono and B.~Williams, ``Iterative risk allocation: A new approach to robust
  model predictive control with a joint chance constraint,'' in {\em IEEE Conf.
  Dec. \& Control}, pp.~3427--3432, 2008.

\bibitem{Zucker2008}
I.~J. Zucker, ``92.34 the cubic equation - a new look at the irreducible
  case,'' {\em The Mathematical Gazette}, vol.~92, no.~524, p.~264–268, 2008.

\bibitem{Gorski2007}
J.~Gorski, F.~Pfeuffer, and K.~Klamroth, ``Biconvex sets and optimization with
  biconvex functions: a survey and extensions,'' {\em Mathematical Methods of
  Operations Research}, vol.~66, pp.~373--407, Dec 2007.

\bibitem{boyd_dc_2016}
T.~{Lipp} and S.~{Boyd}, ``Variations and extension of the convex–concave
  procedure,'' {\em Optimization and Eng.}, vol.~17, pp.~263--287, 2016.

\bibitem{horst2000}
R.~Horst, P.~M. Pardalos, and N.~V. Thoai, {\em Introduction to global
  optimization}.
\newblock Springer Science \& Business Media, 2000.

\bibitem{cvx}
M.~Grant and S.~Boyd, ``{CVX}: Matlab software for disciplined convex
  programming, version 2.1.'' \url{http://cvxr.com/cvx}, Mar. 2014.

\bibitem{gurobi}
L.~Gurobi~Optimization, ``Gurobi optimizer reference manual,'' 2020.

\bibitem{MPT3}
M.~Herceg, M.~Kvasnica, C.~Jones, and M.~Morari, ``{Multi-Parametric Toolbox
  3.0},'' in {\em Proc. Euro. Ctrl. Conf.}, (Z\"urich, Switzerland),
  pp.~502--510, July 17--19 2013.

\bibitem{wiesel1989_spaceflight}
W.~Wiesel, {\em Spaceflight Dynamics}.
\newblock New York: McGraw--Hill, 1989.

\bibitem{Prado2003}
A.~F.~B. de~Almeida~Prado, ``Third-body perturbation in orbits around natural
  satellites,'' {\em Journal of Guidance, Control, and Dynamics}, vol.~26,
  no.~1, pp.~33--40, 2003.

\bibitem{Chihabi2020}
Y.~Chihabi and S.~Ulrich, ``Spacecraft formation guidance law using a state
  transition matrix with gravitational, drag and third-body perturbations,'' in
  {\em Proceedings of the AIAA Scitech 2020 Forum}, AIAA, 2020.

\bibitem{Campi2018TAC}
M.~C. Campi, S.~Garatti, and F.~A. Ramponi, ``A general scenario theory for
  nonconvex optimization and decision making,'' {\em IEEE Trans. Autom.
  Control}, vol.~63, no.~12, pp.~4067--4078, 2018.

\bibitem{Montenbruck2000}
O.~Montenbruck and E.~Gill, {\em Satellite Orbits: Models, Methods and
  Applications}, pp.~53--116.
\newblock Berlin, Heidelberg: Springer Berlin Heidelberg, 2000.

\bibitem{paulson2017stochastic}
J.~Paulson, E.~Buehler, R.~Braatz, and A.~Mesbah, ``Stochastic model predictive
  control with joint chance constraints,'' {\em Int'l J. Ctrl.}, pp.~1--14,
  2017.

\bibitem{Boucheron2013}
S.~Boucheron, G.~Lugosi, and P.~Massart, {\em {Concentration Inequalities: A
  Nonasymptotic Theory of Independence}}.
\newblock Oxford University Press, 02 2013.

\end{thebibliography}

\begin{IEEEbiography}[{\includegraphics[width=1in,height=1.25in,clip,keepaspectratio]{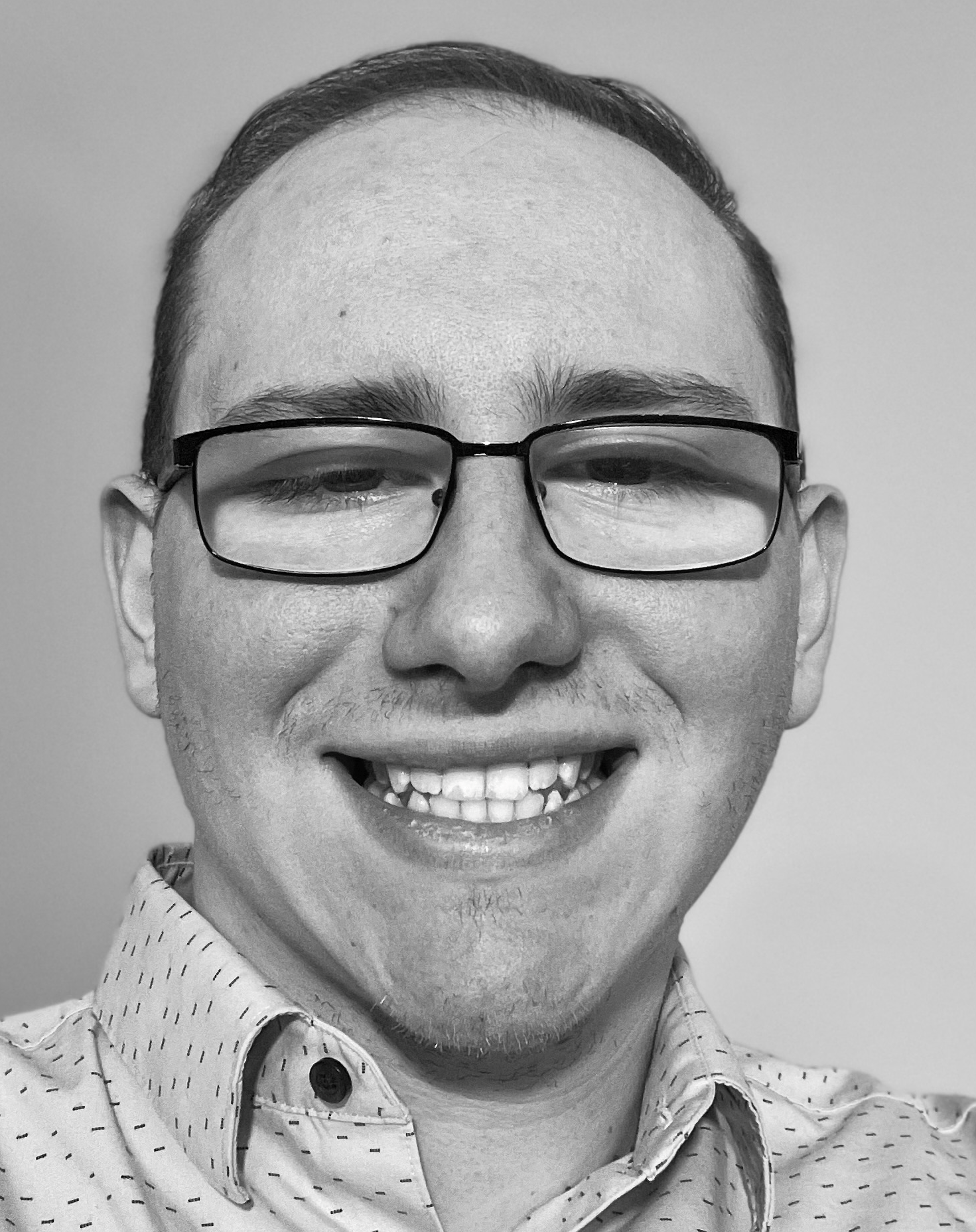}}]{Shawn Priore} (Student Member, IEEE) received the B.S. degree in economics and the M.B.A. degree from John Carroll University, University Heights, OH, USA, in 2016 and 2017, respectively, the M.S. degree in statistics from the University of Connecticut, Storrs, CT, USA. He is currently pursuing the Ph.D. degree in electrical engineering from the University of New Mexico, Albuquerque, NM, USA.

His research interests are in the area of chance constrained stochastic optimal control, autonomous systems, and probabilistic safety, with an emphasis on non-Gaussian disturbances. 

Mr. Priore is the recipient of the Department of Defense SMART Scholarship.
\end{IEEEbiography}

\begin{IEEEbiography}[{\includegraphics[width=1in,height=1.25in,clip,keepaspectratio]{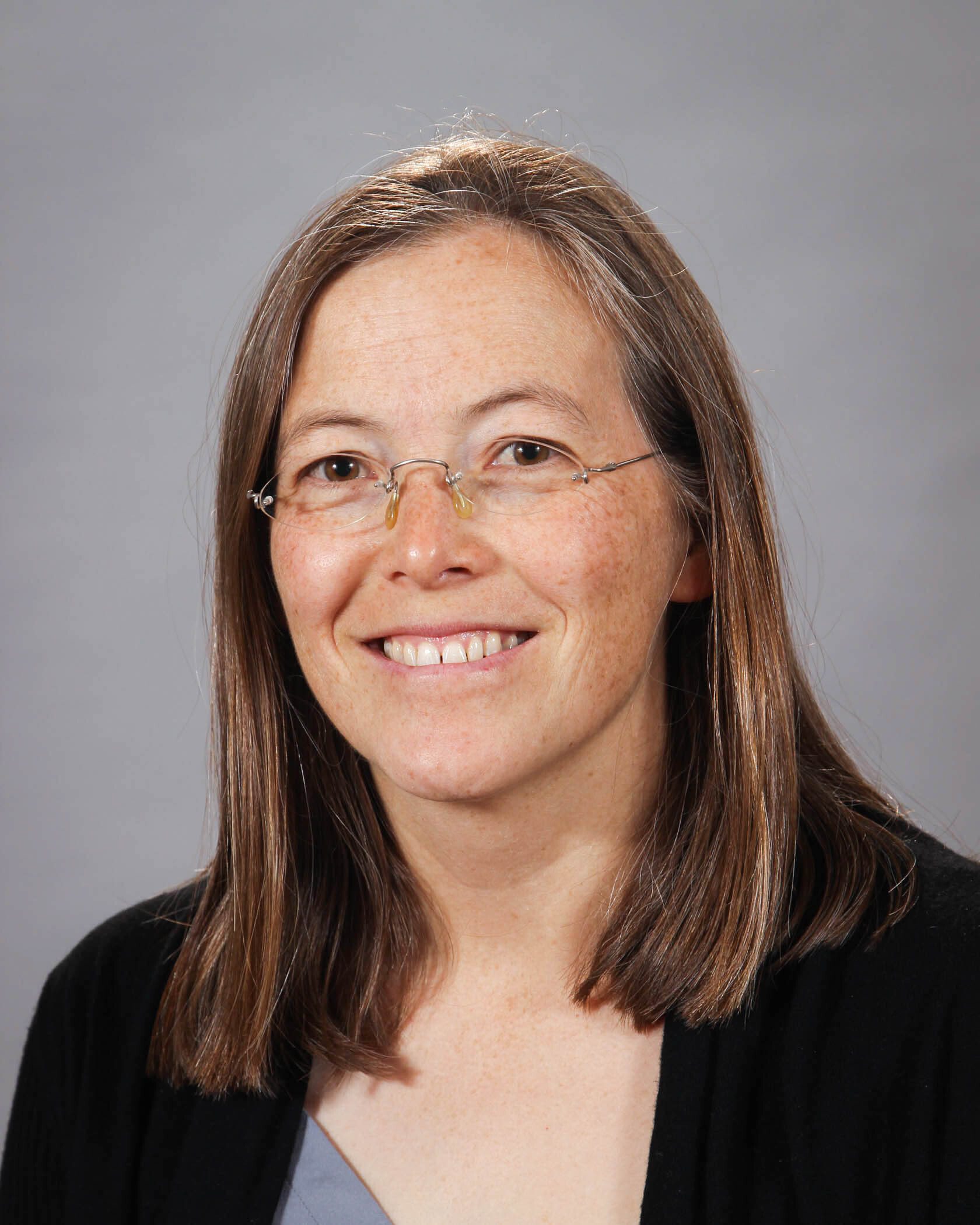}}]{Meeko Oishi} (SM '19, M '04, S '00) received the Ph.D. (2004) and M.S. (2000) in Mechanical Engineering from Stanford University (Ph.D. minor, Electrical Engineering), and a B.S.E. in Mechanical Engineering from Princeton University (1998).  

She is a Professor of Electrical and Computer Engineering at the University of New Mexico.  Her research interests include human-in-the-loop control, stochastic optimal control, and autonomous systems.  She previously held a faculty position at the University of British Columbia at Vancouver, and postdoctoral positions at Sandia National Laboratories and at the National Ecological Observatory Network.  

She is the recipient of the UNM Regents' Lectureship, the NSF CAREER Award, the UNM Teaching Fellowship, the Peter Wall Institute Early Career Scholar Award, the Truman Postdoctoral Fellowship in National Security Science and Engineering, and the George Bienkowski Memorial Prize, Princeton University. She was a Visiting Researcher at AFRL Space Vehicles Directorate, and a Science and Technology Policy Fellow at The National Academies.
\end{IEEEbiography}
\end{document}